\documentclass[11pt]{article}
\usepackage[margin=1in]{geometry}
\usepackage{rpmacros}
\usepackage[T1]{fontenc}
\usepackage[usenames,dvipsnames]{color}
\usepackage{stmaryrd}
\usepackage{lineno}
\usepackage{xfrac}
\usepackage{mathpazo}
\usepackage{todonotes}
\usepackage{setspace}
\usepackage{nicefrac}
\usepackage{gitinfo}
\usepackage{float}
\usepackage{scrextend}
\allowdisplaybreaks

\usepackage{fullpage}
\usepackage[utf8]{inputenc}
\usepackage[russian,english]{babel}

\usepackage{color}
\usepackage[
pagebackref, 
pdfstartview=FitH,pdfpagemode=UseNone,colorlinks=true,citecolor=blue,linkcolor=blue]{hyperref}
\hypersetup{
  linkcolor=[rgb]{0.3,0.3,0.6},
  citecolor=[rgb]{0.2, 0.6, 0.2},
  urlcolor=[rgb]{0.6, 0.2, 0.2}
}

\usepackage[nameinlink,capitalise]{cleveref}
\usepackage{amsthm}
\usepackage{thmtools,thm-restate}

\numberwithin{equation}{section}
\declaretheoremstyle[bodyfont=\it,qed=\qedsymbol]{noproofstyle}

\declaretheorem[numberlike=equation,name=Observation, Refname={Observation,Observations}]{observation}

\declaretheorem[name=Observation,numbered=no]{observation*}

\declaretheorem[numberlike=equation]{theorem}

\declaretheorem[name=Theorem,numbered=no]{theorem*}

\declaretheorem[numberlike=equation]{lemma}
\declaretheorem[name=Lemma,numbered=no]{lemma*}

\declaretheorem[numberlike=equation]{corollary}
\declaretheorem[name=Corollary,numbered=no]{corollary*}

\declaretheorem[numberlike=equation]{proposition}
\declaretheorem[name=Proposition,numbered=no]{proposition*}

\declaretheorem[numberlike=equation,name=Claim, Refname={Claim,Claims}]{claim}
\declaretheorem[name=Claim,numbered=no]{claim*}

\declaretheorem[name=Conjecture,numbered=no]{conjecture*}

\declaretheorem[name=Question,numbered=no]{question*}

\declaretheoremstyle[bodyfont=\it,qed=$\lrcorner$]{defstyle} 

\declaretheorem[numberlike=equation,style=defstyle]{definition}
\declaretheorem[unnumbered,name=Definition,style=defstyle]{definition*}

\declaretheorem[unnumbered,name=Example,style=defstyle]{example*}

\declaretheorem[unnumbered,name=Notation=defstyle]{notation*}

\declaretheorem[unnumbered,name=Construction,style=defstyle]{construction*}

\declaretheorem[numberlike=equation,style=defstyle]{remark}
\declaretheorem[unnumbered,name=Remark,style=defstyle]{remark*}

\usetikzlibrary{decorations.pathreplacing,arrows}

\newif\ifnote
\notefalse
\ifnote
\newcommand{\PHnote}[1]{\textcolor{BrickRed}{\guillemotleft PH: #1\guillemotright}}
\newcommand{\MKnote}[1]{\textcolor{Purple}{\guillemotleft MK: #1\guillemotright}}
\newcommand{\MSnote}[1]{\textcolor{OliveGreen}{\guillemotleft MS: #1\guillemotright}}
\newcommand{\SBnote}[1]{\textcolor{NavyBlue}{\guillemotleft SB: #1\guillemotright}}
\else
\newcommand{\PHnote}[1]{}
\newcommand{\MKnote}[1]{}
\newcommand{\MSnote}[1]{}
\newcommand{\SBnote}[1]{}
\fi

\newcommand{\calC}{\mathcal{C}}
\newcommand{\calL}{\mathcal{L}}
\newcommand{\calT}{\mathcal{T}}
\newcommand{\calG}{\mathcal{G}}
\DeclareMathOperator{\Diag}{Diag}
\DeclareMathOperator{\ord}{ord}
\DeclareMathOperator{\charac}{char}

\newcommand{\ehref}[1]{\href{mailto:#1}{#1}}

\let\epsilon\varepsilon
\newcommand{\coeff}{\operatorname{coeff}}

\newcommand{\enc}{\operatorname{Enc}}

\newcommand{\vu}{\vecu}

\newcommand{\vc}{\vecc}

\onehalfspace

\newcommand{\AFRS}{\text{Additive-FRS}}
\newcommand{\AfFRS}{\text{Affine-FRS}}
\title{Ideal-theoretic Explanation of Capacity-achieving Decoding
\thanks{A conference version of this paper  appeared in the
  \emph{Proc.\ the $25$th International Workshop on Randomization and
    Computation (RANDOM) 2021}~\cite{BhandariHKS2021-affineFRS}.}
}

\author{ 
{Siddharth Bhandari\thanks{Tata Institute of Fundamental Research, Mumbai, India \& Simons Institute for the Theory of Computing, Berkeley, CA, USA. \ehref{siddharth.bhandari@berkeley.edu}. Research partly supported by the Department of Atomic Energy, Government of India, under project no. 12-R\&D-TFR-5.01-0500 and by the Google PhD Fellowship.}}
\and 
{Prahladh Harsha 
\thanks{Tata Institute of Fundamental Research, Mumbai, India. \ehref{prahladh@tifr.res.in}. Research supported by the Department of Atomic Energy, Government of India, under project no. 12-R\&D-TFR-5.01-0500 and in part by  the Swarnajayanti fellowship.}
}
\and
{Mrinal Kumar\thanks{ Department of Computer Science \& Engineering,
    IIT Bombay, Mumbai, India. \ehref{mrinal@cse.iitb.ac.in}. }}
\and
{Madhu Sudan\thanks{School of Engineering and Applied Sciences,
    Harvard University, Cambridge, MA, USA. \ehref{madhu@cs.harvard.edu}. Supported in part by a Simons Investigator Award and NSF Award CCF 1715187.}}
}

\date{}

\begin{document}

\maketitle

\begin{abstract}
  In this work, we present an abstract framework for some algebraic
  error-correcting codes with the aim of capturing codes that are
  list-decodable to capacity, along with their decoding algorithms.  In
  the polynomial ideal framework, a code is specified by some ideals
  in a polynomial ring, messages are polynomials and the encoding of a message polynomial is
  the collection of residues of that polynomial modulo the ideals. We present an alternate way of
  viewing this class of codes in terms of linear operators, and show
  that this alternate view makes their algorithmic list-decodability amenable to analysis.

  Our framework leads to a new class of codes that we call {\em affine Folded Reed-Solomon codes} (which are themselves a special case of the broader class we explore). These codes are common generalizations of the well-studied Folded Reed-Solomon codes and Univariate Multiplicity codes as well as the less-studied Additive Folded Reed-Solomon codes, and lead to a large family of codes that were not previously known/studied.   

  More significantly our framework also captures the algorithmic list-decodability of the constituent codes. Specifically,
  we present
  a unified view of the decoding algorithm for ideal-theoretic codes
  and show that the decodability reduces to the analysis of the
  distance of some related codes. We show that a good bound on this
  distance leads to a capacity-achieving performance of the underlying
  code, providing a unifying explanation of known capacity-achieving
  results.
  In the specific case of affine Folded Reed-Solomon codes, our framework shows that they are efficiently list-decodable up to capacity (for appropriate setting of the parameters), thereby unifying the previous results for Folded Reed-Solomon, Multiplicity and Additive Folded Reed-Solomon codes.

  %        \PHnote{Removing ''and also some known limitations.}
	
%	\MKnote{Didn't follow this line limitations comment: which result is this referring to ?}
\end{abstract}

% \begin{savequuote}
% The search for truth is more precious than its possession.
% \qauthor{---Albert Einstein}
% \end{savequuote}

%%%%%\MSnote{The following is based on an impressionistic view of previous work. Please read carefully and correct.}

\section{Introduction}

Reed-Solomon codes are obtained by evaluations of {polynomials} of degree less than $k$ at $n$ distinct points in a finite field $\F$. Folded Reed-Solomon (FRS) codes are obtained by evaluating a polynomial at $sn$ (carefully chosen) points that are grouped  into $n$ bundles of size $s$ each such that each bundle is in a ge,  and then viewing the resulting $sn$ evaluations as $n$ elements of $\F^s$.
A different grouping of the $sn$ points leads to the less-studied family {of} codes
called the Additive-FRS {codes}.
Multiplicity codes are obtained by evaluating the polynomial and $s-1$ of its derivatives{,} and again viewing the resulting $sn$ evaluations as $n$ elements of $\F^s$. 

This ``bundling'' (or folding, as it is called for FRS codes) in FRS
codes and {multiplicity} codes may be viewed at best as a harmless
operation --- it does not hurt the rate and (relative) distance of the
codes, which {are} already optimal in these parameters. But far from
merely being harmless, in the context {of} algorithmic list-decoding,
bundling has led to remarkable improvements and to two of the very few
explicit capacity achieving codes available in the literature. Indeed, the only
other codes that are known to achieve list-decoding capacity algorithmically and do
not use one of the above codes as an ingredient are the Folded
Algebraic-Geometric codes, which also use bundling. Despite its
central role, the bundling operation is not well-understood
algebraically: it seems like an ``adhoc'' operation rather than
a principled one.  Unearthing what bundling is and  understanding when
and why it turns out to be so powerful is the primary goal of this paper, and we make some progress towards this. 

%\MKnote{Do we really manage to do this here ? We have a framework which captures some of these codes and algorithms, but do we really get any insight on the bundling thing ?}

Turning to the algorithms for list-decoding the above codes close to
capacity, there are two significantly different ones in the
literature. A (later) algorithm due to Guruswami and
Wang~\cite{GuruswamiW2013}\footnote{We note that the Guruswami-Wang
  algorithm is inspired by an idea due to
  Vadhan~\cite[Theorem~5.24]{Vadhan2012} that shows that it suffices
  to interpolate a polynomial $Q$ which is linear in the
  $y$-variables. However, the algorithm from \cite{Vadhan2012} is not
  applicable to our setting since it uses polynomial factorization as
  well as analysis tools that are specific to Reed-Solomon codes. The
  further simplifications developed in \cite{GuruswamiW2013} are key
  to the applicability in our setting.}  which seems more
generalizable, and the original algorithm of Guruswami and
Rudra~\cite{GuruswamiR2008} which is significantly more challenging to
apply to multiplicity codes (see \cite{Kopparty2015}). In both cases,
while the algorithm for FRS works in all (reasonable) settings, the
algorithms for multiplicity codes only work when the characteristic of
the field is larger than the degrees of the polynomials in
question. Looking more closely at FRS codes, part of the careful
choice of bundling in FRS codes is to pick each bundle to be a
geometric progression. If one were to switch this to an arithmetic
progression, then one would get a less-studied family {of} codes
called the Additive-FRS {codes}. It turns out the $\AFRS$ codes are
also known to be list-decodable to capacity but only via the original
algorithm.  We note that the skew polynomial machinery developed by
Gopi and Guruswami~\cite{GopiG2022} in the context of local
reconstruction codes provides yet another proof of list-decodibility
of these codes (See \cref{sec:examples} for more details). Thus, the
short summary of algorithmic list-decoding is that there is no short
summary! {Algorithms tend to work but we need to choose carefully and
  read the fine print.}

The goal of this write-up is to provide a unifying algebraic framework
that (a) captures bundling algebraically, (b) captures most of the
algorithmic success also algebraically, leaving well-defined parts for
combinatorial analysis and (c) leads to new codes that also achieve
capacity. In this work we use basic notions from linear algebra and
polynomial rings to present a unifying definition (see \cref{def: poly
  ideal codes,def: linear op codes}) that captures the codes very
generally, and also their efficient decoding {properties} (see
\cref{thm:intro}). We also describe some new variants of these codes
(see \cref{sec: popular codes as poly ideal codes}), that can be
analyzed using this unifying framework and shown to achieve capacity We
elaborate on these below.

%\MKnote{Again, i am not super clear about this part about bundling here - item (a) above.}
\paragraph{Polynomial ideal codes.} 
Our starting point is what we term ``polynomial ideal codes''.
A \emph{polynomial ideal code} over a finite field $\F$ and
parameters $k,s$ is specified by $n$ pairwise relatively prime monic
polynomials $E_0(X),\dots, E_{n-1}(X) \in \F[X]$ of degree equal to
$s$.\footnote{Here $\F[X]$
	refers to the ring of univariate polynomials in the variable $X$
	over the field $\F$ while $\F_{<k}[X]$ refers to the vector-space of
	polynomials in $\F[X]$ of degree strictly less than $k$.} The encoding maps a message $p \in \F^k$ (interpreted as a
polynomial of degree less than $k$) to $n$ symbols as follows:
\begin{align*}
	\F_{<k}[X]  &\longrightarrow (\F_{<s}[X])^n\\
	p(X) &\longmapsto   \left( p(X) \pmod {E_i(X)}\right)_{i=0}^{n-1}
\end{align*}

The codes described above, Reed-Solomon, FRS, Multiplicity and $\AFRS$, are all examples of polynomial ideal codes. (A rigorous proof can be found in \cref{sec: popular codes as poly ideal codes}). For Reed-Solomon codes, this is folklore knowledge: the evaluation point $a_i$ corresponds to going mod $E_i(X) = (X-a_i)$. {By this we mean that value a polynomial at the evaluation point $a_i$ is the same as the remainder obtained when the polynomial is divided by $E_i(X)=(X-a_i)$.}
For any bundling of the Reed-Solomon codes this follows by taking {the} product of the corresponding polynomials. For multiplicity codes of order $s$, the evaluation of a polynomial and its derivatives at $a_i$ corresponds to going modulo $E_i(X) = (X-a_i)^s$. 

The abstraction of polynomial ideal codes is not new to this work. Indeed Guruswami, Sahai and Sudan~\cite[Appendix A]{GuruswamiSS2000} already proposed these codes as a good abstraction of algebraic codes. Their framework is even more general, in particular they even consider non-polynomial ideals such as in $\Z$. They suggest algorithmic possibilities but do not flesh out the details. 
In this work we show (see \cref{sec: poly ideal johnson bound}) that polynomial ideal codes, as we define them, are indeed list-decodable up to the Johnson
radius. We note that the proof involves some steps not indicated in the previous work but for the most part this confirms the previous thinking. 

The abstraction above also captures ``bundling'' (or folding) nicely - we get {this} by choosing $E_i(X)$ to be a product of some $E_{ij}(X)$.
But the above abstraction thus far fails to capture the capacity-achieving aspects of the codes (i.e., the benefits of this bundling) and the decoding algorithms.
This leads us to the two main \emph{novel} steps of this write-up:
\begin{itemize}
	
	\item We present an alternate viewpoint of polynomial ideal \emph{codes} in terms of
	\emph{linear operators}.
	
	\item We abstract the Guruswami-Wang linear-algebraic
	list-decoding \emph{algorithm} in terms of \emph{linear operators}. 
	
\end{itemize} 

The two sets of ``linear operators'', in the codes and in the decoding algorithm, are not the same. But the linearity of both allows them to interact nicely with each other. We elaborate further below after introducing them.

\paragraph{Linear operator codes.}

In this write-up, a linear operator is an $\F$-linear function $L:\F[X]
\to \F[X]$. A \emph{linear operator code} is characterized by
a family of linear operators $\calL=(L_0,\dots,L_{s-1})$, a set $A = \{a_0, \dots,
a_{n-1}\} \subseteq \F$ of evaluation points and $k$ a degree
parameter such that $k \leq s \cdot n$.  The corresponding linear
operator code, denoted by $LO^{A}_k(\calL)$, is given as follows:
\begin{align*}
	\F_{<k}[X]  &\longrightarrow (\F^s)^n\\
	p(X) &\longmapsto   \left( \calL(p)(a_i) \right)_{i=0}^{n-1}\\
\end{align*}

Linear operator codes easily capture polynomial ideal codes.
For instance, the multiplicity codes are linear operator codes wherein
the linear operators are the successive derivative operators.
But they are also too general --- even if we restrict the operators to
map $\F_{<k}[X]$ to itself, an operator allows $k^2$ degrees of freedom. 

We narrow this broad family by looking {at} subfamilies of linear operators and codes.
The specific subfamily we turn {to} are what we call ``ideal linear operators''. 
We say that linear operators $L_0,\dots,L_{s-1}$ are \emph{ideal linear operators} with 
respect to a set $A$ of evaluation points if
for every $a \in A$, the vector space 
\[I^a(\calL) = \{ p \in \F[X] \mid  \calL(p)(a) = \bar{0} \}\]
is an ideal. (When the set of evaluation points is clear from context, we drop the phrase
``with respect to $A$''.)
Linear operator codes corresponding to ideal linear operators are called \emph{ideal linear
	operator codes} (see \cref{def:lin_op,def: linear op codes} for precise definitions).

%\MKnote{We don't seem to call them ideal linear operators later on. Which name are we preferring ? I like Ideal Linear Operators over Ideal Family Pair of linear operators.}

It is not hard to see that a family of linear operators $\calL=(L_0,\dots,L_{s-1})$ has the
ideal property if it satisfies the following \emph{linearly-extendibility} property: There exists a matrix $M(X) \in \F[X]^{s\times s}$
such that for all $p \in \F[X]$ we have 
\begin{equation*}
	\calL(X\cdot p(X)) = M(X) \cdot \calL(p(X)).
\end{equation*}
This motivates yet another class of linear operators and {codes}:
We say that an operator family $\calL$ is a 
\emph{linearly-extendible linear operator} if such a matrix $M(X)$ exists
and the resulting code is said to be a 
\emph{linearly-extendible linear operator code} (see
\cref{def: lin extendible lin ops,def: linear op
	codes}  for precise definitions).

It turns out that these three definitions of codes ---
polynomial ideal codes, ideal linear operator codes and
linearly-extendible linear operator codes --- are equivalent (see \cref{prop:ilo to pic,prop:pic to ilo,cor: ilo-lelo equiv}). 
%\MSnote{Something is unclear here. ``ideal''ity is wrt a specific set $A$ whereas linearly-extendible is not. So how are we dealing with the missing variable. Do we have a statment of the form ``$\calL$ is  linearly-extendible $\Leftrightarrow \forall A$, $\calL$ is a ideal  operator wrt $A$?'' Or something better than this?} \PHnote{The  equivalence is at the level of codes, not at the level of  operators. One direction is true: ``$\calL$ is  linearly-extendible $\Rightarrow \forall A$, $\calL$ is a ideal  operator wrt $A$}
And while the notion of polynomial ideal codes captures the codes mentioned
thus far naturally, the equivalent notion of linearly-extendible codes 
provides a path to understanding the applicability of the 
linear-algebraic list-decoding algorithm of Guruswami and Wang.

While it is not the case that every linearly-extendible linear operator code (and thus every polynomial ideal code) is amenable to this list-decoding algorithm, it turns out that one can extract a nice sufficient condition on the linear-extendibility 
for the algorithm to be well-defined. This allows us to turn the question of list-decodability into a quantitative one --- how many errors can be corrected. And the linear operator framework now converts this question into analyzing the rank of an associated matrix. 

The sufficient condition we extract is the following: we say that an
operator $L:\F[X] \to \F[X]$ is degree-preserving if $\deg_X(Lf) \leq
\deg_X(f)$ for all $f \in \F[X]$. Observe that any degree-preserving
linear operator when restricted to $\F_{<k}[X]$ can be represented by
an upper-triangular matrix in $\F^{k\times k}$.  A family of linear operators
obtained by repeated iteration, $\calL = (I= L^0, L = L^1, L^2,\dots,
L^{s-1})$ is called an \emph{iterative} family.  We associate with any
degree-preserving family $\calL = (L_0,\ldots,L_{s-1})$  of linear
operators a simple matrix in $\F^{s \times k}$ called $\Diag(\calL)$, whose $i$th row is the
diagonal of $L_i$ and consider the code in $\F^k$ generated by
$\Diag(\calL)$.

% Suppose $L:
% \F[X] \to \F[X]$ is a \emph{degree-preserving} linear
% operator. Consider the \emph{iterative family} of linear operator
% defined as follows: $\calL= (I= L^0, L = L^1, L^2,\dots, L^{s-1})$.

% We say that a linearly-extendible linear operator family $\calL$ is an 
% \emph{iterative family} if there exists an operator $L:\F[X]\to\F[X]$ such that
% $\calL= (I= L^0, L = L^1, L^2,\dots, L^{s-1})$. (Here $L^i$ simply refers to the product of $L$ with itself $i$ times, where we view $L:\F[X]\to\F[X]$ as being specified by a matrix in $\F^{k \times k}$.) 
% \MSnote{Actually - I am not sure how to describe $L^i$. Are we viewing $L$ as a matrix and taking powers? Is $k \times k$ the right dimensions for this? Does this only work for degree preserving linear operators? What are the semantics of taking powers --- is it like composing $L:\F[X]\to\F[Y]$ with $L':\F[Y] \to \F[Z]$? Someone should cleanup and clarify after me here.}
% We also associate a simple matrix in $\F^{s \times k}$ with $\calL = (L_0,\ldots,L_{s-1})$ called $\Diag(\calL)$, whose $i$th column is the diagonal of $L_i$ and consider the code in $\F^k$ generated by $\Diag(\calL)$. 

The following theorem now shows that for any degree-preserving
iterative linearly-extendible operator codes, {the} lower bound on the
distance of $\Diag(\calL)$ yields an upper bound on the list size
obtained by the Guruswami-Wang algorithm, even when the number of
errors approaches {$(1-R)$} where $R$ is the rate of the code.    

\begin{theorem}\label{thm:intro}
	Suppose $\F$ is a field of size $q$ and $L:\F[X]\to \F[X]$ a degree-preserving linear operator and $A$ a set of
	evaluation points such that for $\calL= (L^0, L^1,
	\dots,L^{s-1})$ the corresponding code $\calC$ is a linearly-extendible linear
	operator code. Furthermore, if the matrix $\Diag(\calL) \in \F^{s
		\times k}$ formed by stacking the diagonals of the $s$ linear operators
	as the rows is the generator matrix of a code with  distance
	$1-\frac{\ell}{k}$, then, $\calC$ is code with rate $\frac{k}{sn}$ and relative distance $1 -
\frac{k-1}{sn}$ over an alphabet of size $q^s$, and it is list-decodable up to the distance $1-
	\frac{k}{(s-w+1)n} - \frac1w$ with list size $q^\ell$ for any $1
	\leq w \leq s$.
\end{theorem}

We remark that our actual theorem is more general (see
\cref{lem:general}) where we further separate the role of linear
operators used to build the code {from} those that seed the decoding
algorithm. But it immediately implies \cref{thm:intro} above, which in
turn already suffices to capture the capacity achieving decodability
of FRS, multiplicity and additive-FRS codes. The list-decodability of
multipliciy and additive-FRS codes can be proved using
\cref{thm:intro} by working with the linear operators $L(f(X))=Xf'(X)$
and $L(f(X))= X \cdot ( f(X+\beta) - f(X))$ instead of the more
natural operators $L(f(X))=f'(X)$ and $L(f(X))=f(X+\beta)$. It is to
be noted that using these alternate operators does not change the
underlying codes. However, this is not the approach we follow as we
use the more general \cref{lem:general} to establish the capacity
achieving decodability of the above mentioned codes.  Indeed the
generality of the arguments allows us to capture broader families of
codes uniformly, as described next.

\paragraph{A Common Generalization.}
Our framework leads very naturally to a \emph{new} class of codes that we call
the \emph{Affine Folded Reed-Solomon} (Affine-FRS) codes: 
these are codes defined by ideals of the form
$\prod_{i=0}^{s-1} (X - \ell^{(i)}(a))$ where
$\ell(z) = \alpha z + \beta$ is any linear form and
$\ell^{(i)}(z) = \underbrace{\ell(\ell\dots \ell(z)\dots)}_{i \text{
		times}}$ {denotes the $i$-fold composition of the linear form $\ell(z)$.}
These codes generalize all the previously considered codes: The case 
$\ell(z) = \gamma z$ are the FRS codes, the case
$\ell(z) = z$ are the Multiplicity codes, and the case $\ell(z) = z +
\beta$ are the Additive FRS codes!

\begin{theorem}[Informal statement -- see {\cref{thm:afFRS}}]
\label{thm:informal_afFRS}
  Let $\ell$ be any linear form such
  that either $\ord(\ell) \geq
  k$ or ($\charac(\F) \geq k$ and $\beta \neq 0$) \footnote{$\ord(\ell)$ refers to the smallest positive integer $u$
    such that $\ell^{(u)}(z) = z$.}. Then the
  Affine-FRS codes corresponding to the linear form $\ell$ are
  list-decodable up to capacity.
  \end{theorem}
% We remark that one does not expect to prove list-decoding to capacity
% of AfFRS codes for \emph{all} linear forms till one resolves the
% list-decoding capacity of Reed-Solomon codes. More specifically, the
% case of small characteristic and small order of $\ell$ includes
% multiplicity codes over low-characteristic fields, the
% list-decodability up to capacity of which would impl

% \cref{thm:intro} can be used to show that all the above codes are list-decodable (to capacity by setting $s$ and $w$ appropriately) by the Guruswami-Wang algorithm.
  Previously, even for the special case of the Additive FRS codes,
  list-decodability close to capacity was only achieved by the more
  involved algorithm of \cite{GuruswamiR2008} and
  Kopparty~\cite{Kopparty2015} (see paragraph on Additive Folding and
  Footnote~4 in \cite[Section III]{Guruswami2011}). (A similar approach can be extended to cover the case of $\ord(\ell)\geq k$ in \cref{thm:informal_afFRS}: however, it seems difficult to do so for the case when $\ord(\ell)$ is small.)

  Thus, our Affine-FRS codes lead to
  the first common abstraction of the three codes as well as the first common algorithm for solving the list-decoding problem for these
  codes. (Furthermore, this algorithm is linear-algebraic.)` 
  Arguably thus, even if the Affine-FRS codes had been studied
  previously, it is not clear that the ability to decode them for
  every choice of $\ell(z)$ would be obvious.

\subsection*{Organization}

The rest of the write-up is organized as follows. We begin with some
preliminaries in \cref{sec:prelim}. We then formally define polynomial
ideal codes and linear operator codes in \cref{sec:poly ideal
  codes,sec:linear operators} respectively. In
\cref{sec:listdecoding}, we discuss list-decoding algorithms for
polynomial ideal codes. We first present the list-decoding algorithm
for \emph{all} polynomial ideal codes up to the Johnson radius in
\cref{sec:johnson} and then the list-decoding algorithm beyond the
Johnson radius for special families of linear operator codes in
\cref{sec:general}. The proofs of these algorithms can be found in
\cref{sec: poly ideal johnson bound,sec:list-dec-loc}
respectively. Finally, we conclude by demonstrating how these results can be
used to show that several
well-known families of codes (Folded Reed-Solomon, multiplicity,
additive Folded Reed-Solomon codes) as well as their common
generalization affine
folded Reed-Solomon achieve list-decoding capacity in
\cref{sec:examples}.

\section{Notation \& Preliminaries}\label{sec:prelim}
We start with some notations that we follow in the rest of this write-up. 
\begin{itemize}
\item For a natural number $n$, $[n]$ denotes the set $\{0, 1, \ldots, n-1\}$. 
\item $\F$ denotes a field. 
\item For $a, b, i, j \in \Z$, where $a, b, i, j \geq 0$ the bivariate monomial $X^iY^j$ is said to have $(a, b)$-weighted degree at most $d$ if $ai + bj \leq d$. $N(a, b)$ denotes the number of bivariate monomials of $(1, a)$-weighted degree at most $b$. 
\item For $a, b \in \Z$, a bivariate polynomial $Q(X, Y)$ is said to have $(a, b)$-weighted degree at most $d$, if it is supported on monomials of $(a, b)$-weighted degree at most $d$. 
\item We say that a function $f(n):\N \to \N$ is $\poly(n)$, if there are constants $c, n_0 \in \N$ such that for all $n \geq n_0$, $f(n) \leq n^c$.
\item $\F[X]$ is the ring of univariate polynomials with coefficients in $\F$, and for every $k \in \N$, $\F_{< k}[X]$ denotes the set of polynomials in $\F[X]$ of degree strictly less than $k$.  
\item For a multivariate polynomial $f(X_0, X_1, \ldots, X_{n-1}) \in \F[X_0, X_1, \ldots, X_{n-1}]$, $\deg_{X_i}(f)$ denotes the degree of $f$, when viewing it as a univariate in $X_i$, with coefficients in the polynomial ring on the remaining variables over the field $\F$.%$\F[X_1, \ldots, X_{i-1}, X_{i + 1}, \ldots, X_n]$.
\end{itemize}

\paragraph*{Estimates on number of bivariate monomials:}
We rely on the following simple lemma to estimate the number of bivariate monomials with $(1, a)$-weighted degree at most $b$.
\begin{lemma}\label{lem: wtd deg monomials}
For every $a, b \in \N$, let $N(a,b)$ denote the number of bivariate monomials with $(1, a)$-weighted degree at most $b$. 
Then, the following are true. 
\begin{enumerate}
\item $N(a-1, b) \geq b^2/(2a)$. 
\item For every $\eta \in \N$, if $a$ divides $b$, then 
\[
N(a, b) - N(a, b-a\eta) - \eta(b - a\eta + 1) = a\eta(\eta+1)/2 \, .  
\] 
\end{enumerate} 
\end{lemma}
\begin{proof}
Note that from definition of $N(a, b)$, it follows that $N(a-1, b) \geq N(a, b)$. So to prove the first item of the claim, it suffices to prove a lower bound on $N(a, b)$. 
Let $\tau = \lfloor \frac{b}{a} \rfloor$. Then, 
\begin{align*}
N(a, b) &= \sum_{j = 0}^{\tau} \sum_{i = 0}^{b - aj} 1  \\
&= \sum_{j = 0}^{\tau} (b - aj + 1) \\
&= (b + 1)(\tau + 1) - a \tau (\tau + 1)/2 \\
&= (\tau + 1)/2 \cdot (2b + 2 - a\tau) \, .
\end{align*}
Now, plugging in the value of $\tau$, we get the lower bound on $N(a-1, b)$. 

For the second item, we know that $\tau = b/a$ is an integer. Thus, 
\[
N(a, b) = (b + a)(b + 2)/(2a) \, .
\]
Now, we plug in the exact simplified expression obtained for $N(a, b)$ above in the expression $N(a, b) - N(a, b-a\eta) - \eta(b - a\eta + 1)$ that we aim to estimate, to get the following. 
\begin{align*}
&N(a, b) - N(a, b-a\eta) - \eta(b - a\eta + 1) \\
&=  (b + a)(b + 2)/(2a) - (b - a\eta + a)(b - a\eta + 2)/(2a) - \eta(b - a\eta + 1) \\
&= \frac{1}{2a}\left((b^2 + (2 + a)b + 2a) - ((b - a\eta)^2 + (2 + a)(b - a\eta) + 2a) \right) - \eta(b - a\eta + 1) \\
&= \frac{1}{2a}\left((2 + a)a\eta + 2ab \eta - a^2\eta^2 \right) - \eta(b - a\eta + 1)\\
&= ((1 + a/2)\eta + b\eta - a\eta^2/2) - (b\eta - a\eta^2 + \eta) \\
&= a\eta/2 + a\eta^2/2 \\
&= a\eta(\eta + 1)/2. &\qedhere 
\end{align*}
\end{proof}

\paragraph*{Coding theory basics:}

\begin{definition}[codes, rate, distance]
  Let $\Sigma$ be a finite alphabet and $n$ be a positive
  integer. Given a subset  $C \subseteq \Sigma^n$, define the
  following quantities $R_c$ and $\delta_C$:
  \[ R_C :=\frac{\log_{|\Sigma|}(|C|)}{n}, \qquad \delta_C :=
    \min\limits_{\substack{x,y\in C\\ x\neq
        y}}\left\{\frac{\Delta(x,y)}{n}\right\} \] where $\Delta(x,y)=|\{i\in \{1,2,\ldots,n\}\colon x_i\neq y_i\}|$ denotes the Hamming distance between $x$ and $y$.
  Then, $C$ is said to a \emph{code} of \emph{relative distance}
  $\delta_C$ and \emph{rate} $R_C$ with \emph{blocklength} $n$ over
  the alphabet $\Sigma$.
\end{definition}

\begin{definition}[linear codes] Let $\F_q$ be a field and let $\Sigma = \F_q^s$ for some
  positive integer $s$. We say that $C \subseteq (\Sigma)^n$ is a
  \emph{linear code} if $C$ is an $\F_q$-linear space when viewed as a
  subset of $\F_q^{sn}$.
\end{definition}
We note that the \emph{standard} definition of linear codes
corresponds to the case when $s=1$. All the codes we
consider in this paper will be linear under this slightly more
general definition of linear codes.
The above more general definition allows the alphabet to be a
  linear space $\F_q^s$ instead of just a field $\F_q$. It is easy to
  check that the rate and distance of linear codes
  satisfy $R_C = \frac{\dim_{\F_q}(C)}{sn}$ and $\delta_C=
  \min\limits_{\substack{x\in C\\x\neq 0}}\left\{\frac{|x|}n\right\}$
  where $|x|=|\{i\in \{1,2,\ldots,n\}\colon x_i\neq 0\}|$ denotes the
  Hamming weight of $x \in  (\F_q^s)^n$.

As a consequence of the triangle inequality for Hamming distance, we have that for any code $C\subseteq \Sigma^n$ with relative distance $\delta$ and for all $x\in \Sigma^n$ the number of codewords in the ball of radius $\delta/2$ (in terms of relative distance) centered at $x$, i.e., $\{y\in \Sigma^n\colon \Delta(x,y)/n<\delta/2\}$, is at most $1$. Hence, $\delta/2$ is the so-called \emph{unique-decoding radius} for a code with relative distance $\delta$. A natural question is to ask what happens to the number of codewords within a ball of radius larger than $\delta/2$ centered at $x\in \Sigma^n$. The following well-known fact shows that the number remains polynomial in $n$ even when the radius of the ball grows to $1-\sqrt{(1-\delta)}$.

\begin{theorem}[list-decoding up to Johnson radius {\cite[Theorem 7.3.1]{GuruswamiRS}}]\label{thm: johnson}
Let $q \in \N$ be a natural number. Any code with block length $n$ and relative distance $\delta$ over an alphabet of size $q$ is (combinatorially) list decodable from $(1 - \sqrt{(1-\delta)})$ fraction of errors with list size at most $n^2q\delta$.  
\end{theorem}
% We have the following bound for codes, referred to popularly as the \emph{Singleton bound}~\cite{Singleton1964}, though the bound appears earlier in the works of Joshi~\cite{Joshi1958} and Komamiya~\cite{Komamiya1953}.

% \begin{theorem}[Singleton bound]
% The rate $R$ and the relative distance $\delta$ of a code satisfy $R + \delta \leq 1 + o(1)$. 
% \end{theorem}
% In particular, for codes which lie on the Singleton bound, we have that they are combinatorially list decodable from $1 - \sqrt{R} - o(1)$ fraction errors with polynomial list size. 
We have the following bound for codes, referred to popularly as the \emph{Singleton bound}~\cite{Singleton1964}, though the bound appears earlier in the works of Joshi~\cite{Joshi1958} and Komamiya~\cite{Komamiya1953}.
\begin{theorem}[Komamiya-Joshi-Singleton bound {\cite[Theorem 4.3.1]{GuruswamiRS}}]
The rate $R$ and the relative distance $\delta$ of a code satisfy $R + \delta \leq 1 + o(1)$. 
\end{theorem}
In particular, for codes which lie on the Komamiya-Joshi-Singleton bound, we have that they are combinatorially list decodable from $1 - \sqrt{R} - o(1)$ fraction of errors with polynomial list size. 
\paragraph*{List-decoding upto capacity:}
	
\begin{definition}[list-decoding capacity]
Consider a family of codes $\mathcal{C}=\set{C_1,\ldots,C_n,\ldots}$
where $C_n$ has rate $\rho_n$ and block length $n$ with alphabet
$\Sigma_n$. Then, $\mathcal{C}$ is said to achieve list-decoding
capacity if $\forall \epsilon>0$ there exists an $n_0$ such that
$\forall n\geq n_0$ and all received words $w \in \Sigma^n$, there
exists at most a polynomial number of codewords $c \in C_n$ such that
$\delta(c,w) \leq (1-\rho_n(1+\epsilon))$ where $\delta(c,w)$ is the fractional distance between $c$ and $w$, i.e., $\Delta(c,w)/n$.

Further, if there exists an efficient algorithm for finding all these codewords, then, $\mathcal{C}$ is said to achieve list-decoding capacity efficiently. Ideally, we want to keep $\Sigma_n$ as small as possible.
\end{definition}

\paragraph*{Chinese remainder theorem:}
We also rely on the following version of the Chinese Remainder Theorem for the polynomial ring. 
\begin{theorem}[Chinese remainder theorem {\cite[Corollary 5.3]{GathenG-MCA}}]\label{thm: CRT}
Let $E_0(X), E_1(X), \ldots, E_{n-1}(X)$ be univariate polynomials of degree equal to $s$ over a field $\F$ such that for every distinct $i, j \in [n]$, $E_i$ and $E_j$ are relatively prime. Then, for every $n$-tuple of polynomials $(r_0(X), \ldots, r_{n-1}(X)) \in  \F[X]^n$ such that each $r_i$ is of degree strictly less than $s$ (or zero), there is a unique polynomial $p(X) \in \F[X]$ of degree at most $ns- 1$ such that for all $i \in [n]$,
\[
p(X) = r_i(X) \mod E_i(X) \, .
\] 
\end{theorem}
\paragraph*{Polynomial ideals:}
\begin{definition}\label{def: ideals}
A subset $I \subseteq \F[X]$ of polynomials is said to be an ideal if the following are true.
\begin{itemize}
\item $0 \in I$.
\item For all $p(X), q(X) \in I$, $p + q \in I$. 
\item For every $p(X) \in I$ and $q(X) \in \F[X]$, $p(X)\cdot q(X) \in I$. 
\end{itemize} 
\end{definition}
For the univariate polynomial ring $\F[X]$, we also know that every ideal $I$ is principal, i.e., there exists a polynomial $p(X) \in I$ such that 
\[
I = \{p(X)q(X) : q(X) \in \F[X]\} \, .
\]

\section{Polynomial Ideal Codes}\label{sec:poly ideal codes}
In this section, we discuss polynomial ideal codes in more detail, and
see how this framework captures some of the well studied families of
algebraic error correcting codes.

We start with the formal definition of polynomial ideal codes.  
\begin{definition}[polynomial ideal codes]\label{def: poly ideal codes}
Given a field $\F$, parameters $s, k$ and $n$ satisfying $k < s\cdot n$, the polynomial ideal code
is specified by a family of $n$ polynomials $E_0, \dots,E_{n-1}$ in the ring
$\F[X]$ of univariate polynomials over the field $\F$ satisfying the
following properties.

\begin{enumerate}
\item For all $i \in [n]$, polynomial $E_i$ has degree exactly $s$.
\item The $E_i$'s are \emph{monic} polynomials.
\item The polynomials $E_i$'s are pairwise relatively prime.
\end{enumerate}
The encoding of the polynomial ideal code maps is as follows:
\begin{align*}
  \F_{<k}[X]  &\longrightarrow (\F_{<s}[X])^n\\
  p(X) &\longmapsto   \left( p(X) \pmod {E_i(X)}\right)_{i = 0}^{n-1}
\end{align*}
\end{definition}
As is clear from the definition, polynomial ideal codes are linear
over $\F$ and have rate code $k/(sn)$ and  relative distance $(1 -
(k-1)/(sn))$ (there can't be too many zeros in the encoding of a message polynomial $p(X)$, as the product of all $E_i$'s where the encoding of $p(X)$ is zero, divides $p(X)$). Since the sum of rate and relative distance satisfy the
Singleton bound, these codes are maximal-distance separable (MDS) codes.

We note that in general, $E_i$'s need not have the same degree, but for notational convenience, we work in the setting when each of them is of degree equal to $s$. We also note that these codes continue to be well-defined even if the $E_i$'s are not relatively prime. In this case, the condition, $k < s\cdot n$ is replaced by $k$ being less than the degree of the lowest common multiple of $E_0, E_1, \ldots, E_{n-1}$. However, the distance of the code suffers in this case, and such codes need not approach the Singleton bound. 
%We note that even though we insist on $E_i$s having identical degrees, and them being relatively prime in \autoref{def: poly ideal codes} both these coditions aren't absolutely necessary. In general, we can work with $E_i$'s of varying degrees, and 
%\begin{remark}
%  \begin
%\item The rate of the code is $k/(sn)$ and the distance of the code is
%  $1-(k-1)/(sn)$ (the code is an MDS code). \SBnote{changed distance here}
%  \item The degree of the $E_i$'s need not be identical, but we find
%    it convenient to work with identical degrees.
%    \item The code is well-defined even if the $E_i$'s are not
%      relatively prime in which case the condition $k < s\cdot n$ is
%      replaced by $k < \deg_X (LCM(\{E_i\})$. However, the code is no
%      longer MDS in this case. 
%\end{itemize}
%\end{remark}
We now observe that some of the standard and well studied family of algebraic error correcting codes are in fact instances of polynomial ideal codes for appropriate choice of $E_0, E_1, \ldots, E_{n-1}$. 
\subsection{Some Well Known Codes via Polynomial Ideals}\label{sec: popular codes as poly ideal codes}
The message space for all these codes is identified with univariate polynomials of degree at most $k-1$ in $\F[X]$. We assume that the underlying field $\F$ is of size at least $n$ for this discussion, else, we work over a large enough extension of $\F$. 
\paragraph{Reed-Solomon Codes. } Let $a_0, a_1, \ldots, a_{n-1}$ be $n$ distinct elements of  $\F$.  In a Reed-Solomon code, we encode a message polynomial $p(X) \in \F[X]_{<k}$ by its evaluation on $a_0, a_1, \ldots, a_{n-1}$. To view these as a polynomial ideal code, observe that $p(a_i) = p(X) \mod (X-a_i)$. Thus, we can set the polynomials $E_i(X)$ in \cref{def: poly ideal codes} to be equal to $(X - a_i)$ for each $i \in [n]$. Thus, $s = 1$. Clearly, the $E_i$'s are relatively prime since $a_0, a_1, \ldots, a_{n-1}$ are distinct. 

\paragraph{Folded Reed-Solomon Codes \cite{Krachkovsky2003,GuruswamiR2008}. } Let $\gamma \in \F_q^{*}$ be an element of multiplicative order at least $s$, i.e., $\gamma^{0}, \gamma, \ldots, \gamma^{s-1}$ are all distinct field elements.
Further, let the set of evaluation points be $A=\set{a_0,\ldots,a_{n-1}}$ such that for any two distinct $i$ and $j$ the sets $\set{a_i,a_i\gamma,\ldots,a_i\gamma^{s-1}}$ and $\set{a_j,a_j\gamma,\ldots,a_j\gamma^{s-1}}$ are disjoint.
In a Folded Reed-Solomon code, with block length $n$ and folding parameter $s$ is defined by the following encoding function. 
\[
  p(X) \longmapsto   \left( p(a_i), p(a_i\gamma^{1}), \ldots, p(a_i\gamma^{s-1})\right)_{i = 0}^{n-1}
\]
Thus, these are codes over the alphabet $\F^s$. 

To view these as polynomial ideal codes, we set $E_i(X) = \prod_{j = 0}^{s-1} (X - a_i\gamma^{j} )$. Clearly, each such $E_i$ is a polynomial of degree equal to $s$, and since for any two distinct $i$ and $j$ the sets $\set{a_i,a_i\gamma,\ldots,a_i\gamma^{s-1}}$ and $\set{a_j,a_j\gamma,\ldots,a_j\gamma^{s-1}}$ are disjoint, the polynomials $E_{0}, E_1, \ldots, E_{n-1}$ are all relatively prime.

To see the equivalence between these two viewpoints observe that $p(a_i\gamma^{j} ) = p(X) \mod (X-a_i\gamma^{j} )$. Moreover, $(X-a_i\gamma^{j})$ are all relatively prime as $j$ varies in  $[s]$ for every $i \in [n]$. Thus, by the Chinese Remainder Theorem over $\F[X]$, there is a bijection between remainders of a polynomial modulo $\{(X-a_i\gamma^{j}) : j \in [s]\}$ and the remainder modulo the product $E_i = \prod_{j \in [s]} (X-a_i\gamma^{j} )$ of these polynomials.

\paragraph{Additive Folded Reed-Solomon Codes \cite{GuruswamiR2008}. } Additive Folded Reed-Solomon codes are a variant of the Folded Reed-Solomon codes defined above. 
Let $\F_q$ have characteristic at least $s$ and let $\beta \in \F_q^{*}$.
Further, let the set of evaluation points be $A=\set{a_0,\ldots,a_{n-1}}$ where $a_i-a_j\notin \set{0,\beta,2\beta,\ldots,(s-1)\beta}$ for distinct $i$ and $j$. Here, $s$ denotes the folding parameter. The encoding is defined as follows. 
\[
  p(X) \longmapsto   \left( p(a_i), p(a_i + \beta), \ldots, p(a_i + \beta(s-1))\right)_{i = 0}^{n-1}
\]
Thus, these are also codes over the alphabet $\F^s$. 

To view these as polynomial ideal codes, we set $E_i(X) = \prod_{j = 0}^{s-1} (X - {a_i - \beta j})$. Clearly, each such $E_i$ is a polynomial of degree equal to $s$, and since  $a_i-a_j\notin \set{0,\beta,2\beta,\ldots,(s-1)\beta}$ for distinct $i$ and $j$, the polynomials $E_{0}, E_1, \ldots, E_{n-1}$ are all relatively prime. 

To see the equivalence between the two definitions,  the argument is again identical to that for Folded Reed-Solomon codes discussed earlier in this section. We just observe $(X-{a_i - \beta j})$ are all relatively prime $j$ varies in  $[s]$ for every $i \in [n]$, and thus  by the Chinese Remainder Theorem over $\F[X]$, there is a bijection between remainders of a polynomial modulo $\{(X-{a_i - \beta j} ) : j \in [s]\}$ and the remainder modulo the product $E_i = \prod_{j \in [s]} (X-{a_i - \beta j} )$ of these polynomials.

\paragraph{Univariate Multiplicity Codes \cite{RosenbloomT1997,Nielsen2001,KoppartySY2014}. } Univariate multiplicity codes, or simply multiplicity codes are a variant of Reed-Solomon, where in addition to the evaluation of the message polynomial at every $a_i$, we also give the evaluation of its derivatives of up to order $s-1$. While they can be defined over all fields, for the exposition in this write-up, we consider these codes over fields $\F$ of characteristic at least $sn$. Moreover, we also work with the standard derivatives (from analysis), as opposed to Hasse derivatives which is typically the convention in coding theoretic context. Let $a_0, a_1, \ldots, a_{n-1} \in \F$ be distinct field elements. 

The encoding is defined as follows. 
\[
  p(X) \longmapsto   \left( p(a_i), \frac{\partial p}{\partial X}(a_i), \ldots, \frac{\partial^{s-1} p}{\partial X^{s-1}}(a_i)\right)_{i = 0}^{n-1}
\]
Here, $\frac{\partial^j p}{\partial X^{j-1}}$ denotes the (standard) $j$th order derivative of $p$ with respect to $X$. 

To view these as polynomial ideal codes, we set $E_i(X) = (X - a_i)^s$. Clearly, each such $E_i$ is a polynomial of degree equal to $s$, and since $a_i$'s are all distinct,  these polynomials $E_{0}, E_1, \ldots, E_{n-1}$ are all relatively prime. 

The equivalence of these two definitions follows from an application of Taylor's theorem to univariate polynomials, which says the following.  
\[
p(X) = p(a_i + X -a_i) = p(a_i) + (X-a_i)\frac{\partial p}{\partial X}(a_i) + \cdots + \frac{1}{(s-1)!} (X-a_i)^{s-1} \frac{\partial^{s-1} p}{\partial X^{s-1}}(a_i) + (X-a_i)^s \cdot q(X)\, ,
\]
for some polynomial $q(X) \in \F[X]$. Thus, 
\[
p(X) \mod (X - a_i)^s = p(a_i) + (X-a_i)\frac{\partial p}{\partial X}(a_i) + \cdots + \frac{1}{(s-1)!} (X-a_i)^{s-1} \frac{\partial^{s-1} p}{\partial X^{s-1}}(a_i) .
\]
Therefore, we can \emph{read} off the evaluations of the derivatives of $p$ of order up to $s-1$ at $a_i$ by explicitly writing $p(X) \mod (X - a_i)^s$ as a polynomial in $(X-a_i)$ (via interpolation for instance), and reading off the various coefficients. Similarly, using the above expression, given the evaluation of all the derivatives of order up to $s-1$ of $p$ at $a_i$, we can also reconstruct $p(X) \mod (X - a_i)^s$.

\paragraph{Affine Folded Reed-Solomon Codes}

We now describe a common generalization of the codes defined above, which we call Affine Folded Reed-Solomon Codes.
	Fix integers $k,n,q$ with $n\leq q$. Let $\alpha \in \F_q^*$
        and $\beta\in \F_q$ such that the multiplicative order of $\alpha$ is $u$. 
    	Further, define $\ell(X)=\alpha X+\beta$ and 
    	\[
    	\ell^{(i)}(X)= \underbrace{\ell(\ell\ldots \ell(X))}_{i \text{
    times}} = \alpha^iX +\beta\cdot\sum_{j=0}^{i-1}\alpha^{j} = \alpha_iX + \beta_i.
    \]
    	In fact, if $\alpha\neq 1$, i.e, $u>1$ then, $\beta_u=\beta\cdot\sum_{i=0}^{u-1}\alpha^i=0$, and hence,
        $\ell^{(u)}(X)=\ell^{(0)}(X)$. Let $\ord(\ell)$ denote the smallest positive integer $t$
    such that $\ell^{(t)}(X) = X$. Note that if $\alpha\neq 1$ then $\ord(\ell)=u$.
    	The message space of the Affine Folded Reed-Solomon code of degree $k$ with block length $n$ and folding parameter $s$ is polynomials of degree at most $k-1$ over $\F[X]$, i.e., $\F_{<k}[X]$ where $\F=\F_q$. Let the set of evaluation points be $A=\set{a_0,\ldots,a_{n-1}}$ such that for distinct $i,j$ the sets $\set{\ell^{(0)}(a_i),\ldots,\ell^{(s-1)}(a_i)}$ and $\set{\ell^{(0)}(a_j),\ldots,\ell^{(s-1)}(a_j)}$ are disjoint.
    	
    	The encoding function of Affine Folded Reed-Solomon Codes is given as: (Recall that $t=\ord(\ell)$; let $s=v\cdot t + r$ where $r<t$.)
    	
\begin{align*}   
p(X) \longmapsto  
    \left(\begin{matrix}
        p(\ell^{(0)}(a_i)) & \frac{\partial p}{\partial X}(\ell^{(0)}(a_i)) & \hdots & \frac{\partial^{v-1} p}{\partial X^{v-1}}(\ell^{(0)}(a_i))& \frac{\partial^{v} p}{\partial X^{v}}(\ell^{(0)}(a_i))\\
        \vdots & \vdots & \hdots & \vdots & \vdots\\
        \vdots & \vdots & \hdots & \vdots & \frac{\partial^{v} p}{\partial X^{v}}(\ell^{(r-1)}(a_i)) \\
         p(\ell^{(t-1)}(a_i)) & \frac{\partial p}{\partial X}(\ell^{(t-1)}(a_i)) & \hdots & \frac{\partial^{v-1} p}{\partial X^{v-1}}(\ell^{(t-1)}(a_i)) &  
    \end{matrix}\right)_{i=0}^{n-1}.
\end{align*}	

Thus, these are also codes over the alphabet $\F^s$.
    	
To view these as polynomial ideal codes we set \[E_i(X)=\prod_{j=0}^{s-1}(X-\alpha_ja_i -\beta_j)=\prod _{j=0}^{r-1}(X-\ell^{(j)}(a_i))^{v+1}\cdot\prod_{j=r}^{t-1}(X-\ell^{(j)}(a_i))^v.\]
For the choice of $A$ as above, the polynomials $E_i=E_i(X)$ are pairwise co-prime. Similar to the previous cases of Folded/Additive Reed-Solomon and Multiplicy codes we have a bijection between the remainders of a polynomial modulo $E_i$ and the encoding of the polynomial at $a_i$.

\subsection{An Alternate Definition}
We now discuss an alternate definition of polynomial ideal codes; the advantage being that this definition ties together the polynomials $E_0, E_1, \ldots, E_{n-1}$ into a single bivariate polynomial. This would be useful later on when we discuss the connection between polynomial ideal codes and linear operator codes. 
\begin{definition}[polynomial ideal codes (in terms of bivariate polynomials)]\label{def: poly ideal codes bivariate defn}
Given a field $\F$, parameters $s, k$ and $n$ satisfying
  $k < s\cdot n$, the polynomial ideal code is specified by a
  bivariate polynomial $E(X,Y)$ over the field $\F$ and a set of $n$
  field elements $a_0, a_1 \dots,a_{n-1}$ in $\F$ satisfying the following
  properties.

\begin{enumerate}
\item $\deg_X E(X,Y) = s$.
\item $E(X,Y)$ is a \emph{monic} polynomial in the variable $X$.
\item The polynomials $E(X,a_i)$'s are pairwise relatively prime.
\end{enumerate}
Since $E$ is monic and has (exact) degree $s$ in the variable $X$, any
polynomial $p \in \F[X]$ has the following unique representation.
\[ p(X) = Q^{(p)}(X,Y) \cdot E(X,Y) + R^{(p)}(X,Y) \qquad \text{ where
  } \deg_X(R^{(p)}(X,Y)) < s. \]
The encoding of the polynomial ideal code maps is as follows:
\begin{align*}
  \F_{<k}[X]  &\longrightarrow (\F_{<s}[X])^n\\
  p(X) &\longmapsto   \left( R^{(p)}(X,a_i) \right)_{i = 0}^{n-1}.
\end{align*}
\end{definition}
The equivalence of \cref{def: poly ideal codes,def: poly ideal codes bivariate defn} is not hard to see. We summarize this in the simple observation below. 
\begin{observation}\label{obs: poly ideal codes equivalence}
\cref{def: poly ideal codes,def: poly ideal codes bivariate defn} are equivalent.
\end{observation}
\begin{proof}
Given a code as per \cref{def: poly ideal codes}, we can view this as a code according to \cref{def: poly ideal codes bivariate defn} by picking $n$ distinct $a_0, a_1, \ldots, a_{n-1} \in \F$ (or in a large enough extension of $\F$ of size at least $n$) and use standard Lagrange interpolation to find a bivariate polynomial $E(X, Y)$ such that for every $i \in [n]$,
\[
E(X, a_i) = E_i \, .
\] 
More precisely, we define $E(X, Y)$ as follows.
\[
E(X, Y) := \sum_{i \in [n]} \left(\prod_{j \in [n]\setminus\{i\}}\frac{(Y - a_j)}{(a_i - a_j)} \right) \cdot E_i(X) \, .
\]
Clearly, $E(X, a_i)$'s are relatively prime, and their degree in $X$ equals $s$ and $E(X,Y)$ is monic in $X$. This is because the coefficient of $X^s$ is a polynomial of degree at most $n-1$ which takes the value $1$ at $a_1,\ldots,a_n$, and so has to be the constant $1$.
The equivalence of the encoding function also follows immediately from the definitions. 

The other direction is even simpler. Given a code as per \cref{def: poly ideal codes bivariate defn}, we can view this as a code as per \cref{def: poly ideal codes} by just setting $E_i(X)$ to be equal to $E(X, a_i)$ for every $i \in [n]$. The condition on the degree of $E_i$ and their relative primality follows immediately from the fact that $E(X,Y)$ is monic in $X$ of degree $s$, and $E(X, a_i)$'s are relatively prime. Once again, the encoding map can be seen to be equivalent in both the cases. 
\end{proof} 

From \cref{obs: poly ideal codes equivalence} and the discussion in \cref{sec: popular codes as poly ideal codes}, the Reed-Solomon codes, Folded Reed-Solomon codes, Additive Folded Reed-Solomon codes and Multiplicity codes can also be viewed as polynomial ideal codes as per \cref{def: poly ideal codes bivariate defn}. 

\begin{itemize}
    \item \textbf{Reed-Solomon codes}: We take $E(X,Y)$ to be equal to $(X-Y)$, the set of points $a_0, \ldots, a_{n-1}$ remain the same.
    \item \textbf{Folded Reed-Solomon codes}: We take $E(X,Y) = \prod_{j\in [s]} (X - \gamma^jY)$ and the set of evaluation points $a_0, \ldots, a_{n-1}$ are set as before, and $\gamma \in \F^{*}$ is an element of high enough order.
    \item  \textbf{Additive Folded Reed-Solomon codes}:  We take $E(X,Y) = \prod_{j\in [s]} (X - Y + \beta j)$ and the set of evaluation points $a_0, \ldots, a_{n-1}$ are set as before. Recall that $\F$ is taken to be a field of characteristic at least $s$ for these codes.
    \item \textbf{Multiplicity codes}: We take $E(X,Y)$ to be equal to $(X-Y)^s$, the set of points $a_0, \ldots, a_{n-1}$ are distinct.
    \item \textbf{Affine Folded Reed-Solomon codes:} We take $E(X,Y)=\prod_{i=0}^{s-1}(X-\ell^{(i)}(Y))$ where $\ell(Y)=\alpha Y+\beta$ with $\alpha\in \F_q^*$ and $\beta\in \F_q$. Recall that the set of evaluation points  $A=\set{a_0,\ldots,a_{n-1}}$ is such that for distinct $i,j$ the sets $\set{\ell^{(0)}(a_i),\ldots,\ell^{(s-1)}(a_i)}$ and $\set{\ell^{(0)}(a_j),\ldots,\ell^{(s-1)}(a_j)}$ are disjoint.
\end{itemize}
It follows immediately from these definitions that all the desired properties in \cref{def: poly ideal codes bivariate defn} are indeed satisfied. We skip the remaining details.   

\section{Linear Operator Codes}\label{sec:linear operators}

In this section, we give an alternate viewpoint of polynomial ideal codes in terms of codes defined based on linear operators on the ring of
polynomials.

\begin{definition}[linear operators]\label{def:lin_op}
  Let $\calL = (L_0,\dots, L_{s-1})$ be a tuple of $s$ linear operators
  where each $L_i:\F[X] \to \F[X]$ is a $\F$-linear operator over the
  ring $\F$.  For any $f \in \F[X]$, it will be convenient to denote by
  $\calL(f)$ the (row) vector
  $(L_0(f), \dots, L_{s-1}(f)) \in \F[X]^s$.

  Given any such family $\calL$ and element $a \in \F$, define
\[ I^a(\calL) = \{ p(X) \in \F[X] \mid  \calL(p)(a) = \bar{0} \}.\]
If the family $\calL$ of linear operators family and the set of
  field elements $A \subseteq \F$ further satisfy the property that
  $I^a(\calL)$ is an ideal for each $a \in A$, we refer to the family
  $\calL$ as an \emph{ideal family of linear
    operators} with respect to $A$.

  In this case, since $\F[X]$ is a principal ideal domain, for each $a\in A$, $I^a(\calL) = \langle E^a(\calL)(X) \rangle$ for some
  monic polynomial $E^a(\calL)\in F[X]$. \qedhere
\end{definition}

We now define a special condition on the family of linear operators $\calL$ which will help us capture when $I^a(\calL)$ forms an ideal.

\begin{definition}[linearly-extendible linear operators]\label{def: lin extendible lin ops}
The family $\calL$ of linear operators is said to be \emph{linearly-extendible} if there exists a matrix $M(X) \in \F[X]^{s\times s}$ such that for all $p \in F[X]$ we have 
  \begin{equation}\label{eq:linear_ext}
    \calL(X\cdot p(X)) = M(X) \cdot \calL(p(X)).
    \end{equation}
\end{definition}

We give two examples to illustrate the definition:
\begin{itemize}
    \item Let $L_0(f(X))=f(X)$ and $L_1(f(X))=f'(X)$ where $f'$ is the formal derivative of $f$. Then, by the product rule $L_1(Xf(X))=X\cdot f'(X)+f(X)$. Hence, in this case $M(X)=\big(\begin{smallmatrix}
  X & 0\\
  1 & X
\end{smallmatrix}\big)$. 

    \item Let $L_0(f(X))=f(X)$ and $L_1(f(X))=f(\gamma X)$ where $\gamma \in \F_q$ is non-zero. Then, we have $L_1(Xf(X))=\gamma Xf(\gamma X)$. Hence, in this case $M(X)=\big(\begin{smallmatrix}
  X & 0\\
  0 & \gamma X
\end{smallmatrix}\big)$. 
\end{itemize}

\begin{observation} Suppose $\calL$ is linearly-extendible and $M(X)$ is the
  corresponding matrix from \cref{eq:linear_ext}.
  \begin{itemize}
  \item For any $j\geq0$ we have $\calL(X^j\cdot p(X))=(M(X))^j\cdot \calL(p(X))$. Thus, by linearity we have that for any $q \in \F[X]$:
  \[ \calL( q(X) \cdot p(X) ) = q(M(X)) \cdot \calL(p(X)).\] For
  instance if $q(X)=X^j$ then $\calL(X^j\cdot p(X))=(M(X))^j\cdot \calL(p(X))$.
  
\item The family $\calL$ is
  completely specified by $\calL(1)$ and $M(X)$. In other words,
  $\calL(p(X))=p(M(X))\cdot \calL(1)$.

\item For every set $A$ of evaluation points, $\calL$ is an ideal
  family of linear operators with respect to $A$. This is because if
  at a point $a$ we have $\calL(p)(a)=0$ then
  $\calL(Xp)(a)=(M(X)\cdot\calL(p(X)))(a)=M(X=a)\cdot \calL(p)(a)=0$
  . This means that if $p(X)\in I^a(\calL)$ then $Xp(X)\in
  I^a(\calL)$, and hence by linearity for any $q(X)\in \F[X]$ we have
  $q(X)\cdot p(X)\in I^a(\calL)$.
\end{itemize}
\end{observation}

\begin{definition}[linear operator codes]\label{def: linear op codes}
Let $\calL=(L_0,\dots,L_{s-1})$ be a family of linear operators, $A = \{a_1, \dots,
a_n\} \subseteq \F$ be a set of evaluation points and $k$ a degree
parameter such that $k \leq s \cdot n$. Then the linear operator
code generated by $\calL$ and $A$, denoted by $LO^{A}_k(\calL)$ is given
as follows:
\begin{align*}
  \F_{<k}[X]  &\longrightarrow (\F^s)^n\\
  p(X) &\longmapsto   \left( \calL(p)(a_i) \right)_{i=1}^n\\
\end{align*}
\begin{itemize}
  \item If $\calL$ is an ideal family of linear operators with respect
    to $A$ where the polynomials $E_i := E^{a_i}(\calL)$, which are the monic generator polynomials for the ideals $I^{a_i}(\calL)$, further satisfy the following:
    \begin{enumerate}
  \item For all $i \in [n]$, polynomial $E_i$ has degree exactly $s$.
  \item The polynomials $E_i$'s are pairwise relatively prime.
  \end{enumerate}
Then the linear operator code is said to be an \emph{ideal linear
  operator code} and denoted by $ILO^A_k(\calL)$.
\item If the ideal linear operator code $ILO^{A}_k(\calL)$ further
  satisfies that $\calL$ is linearly-extendible, then the ideal linear
  operator code is said to be a \emph{linearly-extendible linear
    operator code}, denoted by $LELO^{A}_k(\calL)$.
\end{itemize}
\end{definition}

\begin{remark} The rate of the $LO^{A}_k(\calL)$ code is $k/(sn)$. Further, if the the code is an ideal linear operator code, i.e., $ILO^A_k(\calL)$, then its distance is $1-\frac{k-1}{sn}$. This is because for any message polynomial $p(X)$, the product of all $E_i$'s where the encoding of $p(X)$ is zero, divides $p(X)$, and hence there can't be too many zeros in the encoding of $p(X)$.
Hence, $ILO^A_k(\calL)$ is an $MDS$ code. 
  \end{remark}

\begin{proposition} 
\label{prop:pic to ilo}
Any polynomial ideal code is a linearly-extendible  linear operator code.
\end{proposition}
\begin{proof}
Consider a polynomial ideal code given by a bivariate polynomial $E(X,Y)$ and a set of evaluation points $\set{a_1,\ldots,a_{n}}$ as in \cref{def: poly ideal codes bivariate defn}. Recall that $E(X,Y)$ is a monic polynomial in the variable $X$, $\deg_X E(X,Y)=s$ and the $E(X,a_i)$'s are relatively prime.
Further, any polynomial $p(X) \in \F[X]$ has the following unique representation.
\[ p(X) = Q^{(p)}(X,Y) \cdot E(X,Y) + R^{(p)}(X,Y) \qquad \text{ where
  } \deg_X(R^{(p)}(X,Y)) < s. \]
The encoding map of the polynomial ideal code is as follows:
\begin{align*}
  \F_{<k}[X]  &\longrightarrow (\F_{<s}[X])^n\\
  p(X) &\longmapsto   \left( R^{(p)}(X,a_i) \right)_{i = 0}^{n-1}.
\end{align*}

Let $E(X,Y)=X^s-\sum_{i=0}^{s-1}H_i(Y)X^i$ and
$R^{(p)}(X,Y)=\sum_{i=0}^{s-1}R^p_i(Y)X^i$. Define $\calL=
(L_0,\ldots, L_{s-1})$ as $L_i(p(X))=R^p_i(X)$\footnote{Note that
  we have changed the formal variable from $Y$ to $X$ in the
  definition of $R^p_i(X)$ here.}.
Therefore, at any point $a\in \set{a_1,\ldots, a_n}$ we have $R^{(p)}(X,a)=\sum_{i=0}^{s-1}L_i(p(X))(a)\cdot X^i$.

Notice, that for $p(X),q(X)\in \F[X]$ we have $R^{(p+q)}(X,Y) = R^{(p)}(X,Y)+R^{(q)}(X,Y)$ and thus $R^{p+q}_i(Y)=R^{p}_i(Y)+R^{q}_i(Y)$ for $i<s$. This shows that $L_i$'s are indeed linear operators. 
Also, $R^{(Xp)}(X,Y)=\sum_{i=1}^{s-1}R^p_{i-1}(Y) X^{i}+R^p_{s-1}(Y)\cdot\sum_{i=0}^{s-1}H_i(Y)X^i$. Therefore, we have $\calL(Xp(X))=M(X)\calL(p(X))$ where $M(X)_{ij}=\mathbb{I}[i-1=j]+\mathbb{I}[j=s-1]\cdot H_i(X)$ for $i,j\in \set{0,1,\ldots,s-1}$. More descriptively,

\begin{align*}
 M_{s\times s}=   \begin{bmatrix}
0 & 0 & 0& \ldots & H_0(X) \\
1 & 0 & 0& \ldots & H_1(X)\\ 
0 & 1 & 0 & \ldots & H_2(X)\\
\vdots & \vdots & 1 &\ldots & H_3(X)\\
\vdots & \vdots & \vdots &\ldots &\vdots \\
0 & 0& \ldots & 1 & H_{s-1}(X) 
\end{bmatrix}
\end{align*}

Hence $\calL$ forms a linearly-extendible set of linear operators. 
\end{proof}

\begin{remark}(degree preserving)
  If the bivariate polynomial $E(X,Y)$ has total degree $s$, then, the linear operator in the $LELO$ code obtained above has the property that $\deg_X L_i(X^j) \leq j$: in fact, $\deg_X L_i(X^j) \leq j-i$.
\end{remark}
   
\begin{proposition} \label{prop:ilo to pic}
Any ideal linear operator code is a polynomial ideal code.
\end{proposition}
\begin{proof}
Consider an ideal linear operator code $ILO^A_k(\calL)$. For any polynomial $p(X)\in \F[X]$ and a point $a_i\in A$, giving $\calL(p(X))(a_i)$ is equivalent to giving $p(X) \mod \langle E_i \rangle$ where $\langle E_i \rangle=I^{a_i}(\calL)$.
However, the $E_i$s readily satisfy \cref{def: poly ideal codes}.
\end{proof}

Now, we state a corollary which further corroborates the notion of linear-extendibility.
\begin{corollary}[Equivalence of $ILO$ and $LELO$]\label{cor: ilo-lelo equiv}
From \cref{prop:pic to ilo,prop:ilo to pic} it follows that every ideal linear operator code is also a linearly-extendible linear operator code.
\end{corollary}
\SBnote{define Diag(G) and degree preserving in main-lemma}

\SBnote{Need to mention distance of the code in sec 2. That is they are MDS}

\SBnote{clean up section for i,j starting points and x,y vs X,Y}

\SBnote{definitions from sec 6 to be lifted here}

Below we state some well known codes in their linear operator descriptions (a more formal treatment is given in \cref{sec:examples}):
\begin{itemize}
    \item \textbf{Reed-Solomon Codes}: Let $A=\set{a_0,\ldots,a_{n-1}}$ be distinct elements in $\F_q$
    These are $LELO_{\calL,A}$ where $\calL=(I)$. That is the encoding of the message polynomial $p(X)\in \F_{<k}[X]$ at a point $a$ is  $L(f(X))(a)=f(a)$.
    \item \textbf{Folded Reed-Solomon Codes}: Let $\gamma\in F_q^{*}$ with multiplicative order at least $s$. $FRS[k,n]$ with folding parameter $s$ are linearly-extendible linear operator codes $LELO_{\calL,A}$ where:
	\begin{itemize}
	    \item $\mathcal{L}=(L_0,\ldots,L_{s-1})$ with {$L_1(f(X))= f(\gamma X)$}  for $f(X)\in \F_q[X]$ and $L_i=L_1^i$ for $i \in \set{0,1,\ldots,s-1}$.  
	    \item For the above family of operators $M(X)$ is given by $M(X)_{ij}=\gamma^{i}X\cdot \mathbb{I}[i=j]$ for $i,j\in [s]$.
	    \item The set of evaluation points is $A=\set{a_0,\ldots,a_{n-1}}$ where for any two distinct $i$ and $j$ the sets $\set{a_i,a_i\gamma,\ldots,a_i\gamma^{s-1}}$ and $\set{a_j,a_j\gamma,\ldots,a_j\gamma^{s-1}}$ are disjoint.
	\end{itemize}
	\item \textbf{Multiplicity Codes}:
	 Then, $MULT[k,n]$ codes of order $s$ are linearly-extendible linear operator codes $LELO_{\calL,A}$ where:
	\begin{itemize}
	    \item $\mathcal{L}=(L_0,\ldots,L_{s-1})$ with $L_1(f(X))= \frac{\partial f(X)}{\partial X}$  for $f(X)\in \F_q[X]$ and $L_i=L_1^i$ for $i \in \set{0,1,\ldots,s-1}$.  
	    \item For the above family of operators $M(X)$ is given by $M(X)_{ij}=X\cdot \mathbb{I}[i=j] + i\cdot \mathbb{I}[i-1=j]$ for $i,j\in [s]$.
	    \item The set of evaluation points is $A=\set{a_0,\ldots,a_{n-1}}$ where $a_i$s are all distinct.
	\end{itemize}
	\item \textbf{Additive Folded Reed-Solomon Codes}:
	Let $\beta\in \F_q$ be a non-zero element and the characteristic of $\F_q$ be at least $s$. Then, $\AFRS[k,n]$ codes with folding parameter $s$ are linearly-extendible linear operator codes $LELO_{\calL,A}$ where:
	\begin{itemize}
	    \item $\mathcal{L}=(L_0,\ldots,L_{s-1})$ with $L_1(f(X))= f(X+\beta)$  for $f(X)\in \F_q[X]$ and $L_i=L_1^i$ for $i \in \set{0,1,\ldots,s-1}$.  
	    \item For the above family of operators $M(X)$ is given by $M(X)_{ij}=(X+i\beta)\cdot \mathbb{I}[i=j]$ for $i,j\in [s]$.
	    \item The set of evaluation points is $A=\set{a_0,\ldots,a_{n-1}}$ where $a_i-a_j\notin \set{0,\beta,2\beta,\ldots,(s-1)\beta}$ for distinct $i$ and $j$.
	\end{itemize}
	\item \textbf{Affine Folded Reed-Solomon Codes:}
	Let $\alpha \in \F_q^{*}$ and $\beta \in \F_q$. Further, let $\ell(X)=\alpha X+ \beta$ with $\ord(\ell)=u$. Then $\text{Affine-FRS}[k,n]$ codes with folding parameter $s$ are linearly-extendible codes $LELO_{\calL,A}$ described below.
(See \cref{obs:affrs_lin_op_rep} for more details.)
	
	Define $D_1:\F[X]\to \F[X]$ as $D_1(f(X))=\frac{\partial f(X)}{\partial X}$ and $S_1:\F[X]\to \F[X]$ as $S_1(f(X))=f(\ell(X))$. Further, for $i\geq 0$ let $D_i=D_1^i$ and $S_i=S_1^i$. Recall, that the order of $\alpha$ is $u$. For any integer $r\in[s]$ let $r=r_1u+r_0$, with $r_0<u$, be the unique representation of $r$. 
	\begin{itemize}
	    \item Define $L_{r}:\F[X]\to \F[X]$ as $L_{r}(f(X))=S_{r_0}(D_{r_1}f(X))$. Set $\calL=(L_0,\ldots,L_{s-1})$. Clearly, $\calL$ is a family of linear operators. 
	    \item $L_{r}(Xf)=S_{r_0}(D_{r_1}Xf)=S_{r_0}(r_1\cdot D_{r_1-1}f+X\cdot D_{r_1}f)=r_1\cdot L_{r-u}f+S_{r_0}(X)\cdot L_rf$: hence, $\calL$ is a set of linearly-extendible linear operators.
	    \item The set of evaluation points  $A=\set{a_0,\ldots,a_{n-1}}$ is such that for distinct $i,j$ the sets $\set{\ell^{(0)}(a_i),\ldots,\ell^{(s-1)}(a_i)}$ and $\set{\ell^{(0)}(a_j),\ldots,\ell^{(s-1)}(a_j)}$ are disjoint.
	\end{itemize}

\end{itemize}

\section{List-Decoding of Polynomial Ideal Codes}\label{sec:listdecoding}

% We also discuss a list-decoding
                                % algorithm for all such codes, up to
                                % the Johnson radius. The algorithm is
                                % an adaptation of the algorithm of
                                % Guruswami, Sahai and Sudan
                                % \cite{GuruswamiSS2000} for list
                                % decoding error correcting codes
                                % based on the Chinese Remainder
                                % Theorem (which also have a natural
                                % ideal theoretic view over the
                                % integers) to our setting, with a few
                                % more observations which seem to be
                                % needed which working with polynomial
                                % ideals.

In this section, we discuss the list-decoding of polynomial ideal
codes.

\subsection{List-decoding Up to the Johnson Radius}\label{sec:johnson}

We first observe that polynomial ideal codes are list decodable in polynomial time, up to the Johnson radius. 

\begin{theorem}\label{thm: decoding up to Johnson}
Let $k, s, n \in \N$ be such that $k < sn$ and $s < k-1$. Let $E_0(X), E_1(X), \ldots, E_{n-1}(X) \in \F[X]$ be relatively prime monic polynomials of degree equal to $s$ each. Let $\enc:\F_{<k}[X] \longrightarrow (\F_{<s}[X])^n$ be the encoding function defined as 
\[
p(X) \longmapsto   \left( p(X) \pmod {E_i(X)}\right)_{i = 0}^{n-1} \, .
\]
Then, there is an algorithm, which takes as input a received word $\vc = (\vc_0, \vc_1, \ldots, \vc_n) \in \F_{<s}[X]^n$ and for every $\epsilon > 0$ outputs all polynomials $f \in \F_{<k}[X]$ such that $\enc(f)$ and $\vc$ agree on  at least  $(k/(sn))^{1/2} + \epsilon$ fraction of coordinates  in time $\poly(n, 1/\epsilon)$.    
\end{theorem}

Observe that the rate of this code is $k/(sn)$ and distance is $1 - (k-1)/(sn)$, and thus \cref{thm: decoding up to Johnson} gives us an algorithmic analog of \cref{thm: johnson} for these codes. 

The list-decoding algorithm for polynomial ideal codes is an (almost immediate) extension of an algorithm of Guruswami, Sahai and Sudan \cite{GuruswamiSS2000} for list-decoding codes based on Chinese Remainder Theorem to this setting. This algorithm, in turn, relies on ideas in an earlier algorithm of Guruswami and Sudan \cite{GuruswamiS1999} for list-decoding Reed-Solomon codes up to the Johnson radius. 

As noted in the introduction, most of the ideas for the proof of
\cref{thm: decoding up to Johnson} were already there in the work of
Guruswami, Sahai and Sudan \cite{GuruswamiSS2000} and all we do in
this section is to flush out some of the details. The proof of this
theorem is deferred to \cref{sec: poly ideal johnson bound}.

\subsection{List-decoding beyond the Johnson Radius}\label{sec:general}

In this section, we use the linear operator viewpoint of polynomial
ideal codes to study their list-decodability beyond the Johnson
radius.  We show that if the family of linear operators $\calL$ and
the evaluation points satisfy some further properties, then the linear
operator code is list-decodable all the way up to the distance of the
code.

Let $\calG = (G_0,\dots, G_{w-1})$ and
$\calT = (T_0,T_1,\dots,T_{r-1})$ be two families of linear operators
such that $G_i: \F[X] \to \F[X]$ and $\calT$ is a linearly-extendible
family of linear operators.
  We say that the pair $(\calT,\calG)$ \emph{list-composes} in terms
  of $\calL$ at the set of
  evaluation points $A$ if we have
  the following. For every linear operator $G \in \calG$ and field
  element $a \in A$, there exists a linear function $h_{G,a}: \F^s \to \F^r$
  such that for every polynomial $f \in \F[X]$ we have
\[
  \calT(G(f))(a) = h_{G,a}(\calL(f)(a)).
\]

{For instance, consider the $FRS$ code over $\F_q$ with folding parameter $s$ and with the set of evaluation points being $A=\set{a_0,\ldots,a_{n-1}}$. The message space is polynomials of degree at most $k-1$ over $\F_q[X]$.  This code is a linearly-extendible linear operator code where $\mathcal{L}=(L_0,\ldots,L_{s-1})$ with $L_1(f(X))= f(\gamma X)$  for $f(X)\in \F_q[X]$ and $L_i=L_1^i$ for $i \in \set{0,1,\ldots,s-1}$.  
Set $\mathcal{G}=(L_0,\ldots,L_{w-1})$ for some integer $w<s$ and $\calT=(T_0,\ldots,T_{r-1})$ with $r=s-w+1$ and $T_i=L_i$.
Then, for all $G_i \in \calG$, $T_j\in \calT$ and $a \in A$, we have that for every polynomial $f \in \F[X]$: $T_j(G_i(f))(a) = L_{i+j}(f)(a)$. Notice that $L_{i+j}\in \cal L$ as $i+j \leq s-1$. Hence, the pair $(\calT,\calG)$ \emph{list-composes} in terms of $\calL$ at the set of evaluation points $A$.}
	
% 	Fix integers $k,n,q$ with $n\leq q$. Fix $\gamma\in \F_q^{*}$
%         of multiplicative order at least $s$. The message space of the $FRS^{\gamma}_s[k,n]$ code with folding parameter $s$ is polynomials of degree at most $k-1$ over $\F[X]$, i.e., $\F_{<k}[X]$ where $\F=\F_q$.  Then, $FRS$ codes are linearly-extendible linear operator codes $LELO_{\calL,A}$ where:
% 	\begin{itemize}
% 	    \item $\mathcal{L}=(L_0,\ldots,L_{s-1})$ with $L_1(f(X))= f(\gamma X)$  for $f(X)\in \F_q[X]$ and $L_i=L_1^i$ for $i \in \set{0,1,\ldots,s-1}$.  
% 	    \item For the above family of operators $M(X)$ is given by $M(X)_{ij}=\gamma^{i}X\cdot \mathbb{I}[i=j]$ for $i,j\in [s]$.
% 	    \item The set of evaluation points is $A=\set{a_0,\ldots,a_{n-1}}$ where for any two distinct $i$ and $j$ the sets $\set{a_i,a_i\gamma,\ldots,a_i\gamma^{s-1}}$ and $\set{a_j,a_j\gamma,\ldots,a_j\gamma^{s-1}}$ are disjoint.
% 	\end{itemize}
% We will prove this by applying \cref{lem:general}. Set $\mathcal{G}=(L_0,\ldots,L_{w-1})$ for some integer $w<s$ to be set later and $\calT=(T_0,\ldots,T_{r-1})$ with $r=s-w+1$ and $T_i=L_i$. 

% For all $G_i \in \calG$, $T_j\in \calT$ and $a \in A$, we have that for every polynomial $f \in \F[X]$: $T_j(G_i(f))(a) = L_{i+j}(f)(a)$. Notice that $L_{i+j}\in \cal L$ as $i+j \leq s-1$.

% \SBnote{Should we say that thus function is linear by default?}
\begin{theorem}
\label{lem:general}
  If $LO^A_k(\calL)$ is a linear operator code and there exists two
  families of linear operators $\calG= (G_0,\dots, G_{w-1})$ and $\calT =
  (T_0,\dots,T_{r-1})$ such that
  \begin{enumerate}
  \item $(\calT,A)$ forms a linearly-extendible linear operator code $LELO^A_{k+nr/w}(\calT)$ \label{itm:general_lem_lin_ext}
    \item The pair $(\calT,\calG)$ list-composes in terms of $\calL$ at the set of evaluation
      points \label{itm:general_lem_list_comp}
    \item $\calG$ is degree-preserving \label{itm:general_lem_deg_pre}
      \item $\Diag(\calG) \in \F^{|\calG| \times k}$ is the generator
        matrix of a code with distance $k-\ell$. \label{itm:general_lemma_G_dist}

    \end{enumerate}
    Then, $LO^{A}_k(\calL)$ is list-decodable up to the distance $1-
    \frac{k}{rn} - \frac1w$ with list size $q^\ell$.
    \SBnote{should this change??}
  \end{theorem}

This theorem clearly implies \cref{thm:intro}. Recall the hypothesis of \cref{thm:intro}. We instantiate $\calG, \calT$ in \cref{lem:general} as $(L_0,\dots, L_{w-1})$ and $(L_0,\dots,L_{r-1})$ respectively, with $r=s-w+1$: properties $1,3$ and $4$ above follow directly from the hypothesis of \cref{thm:intro}. For property $2$ notice that for all $G_i \in \calG$, $T_j\in \calT$ and $a \in A$, we have that for every polynomial $f \in \F[X]$: $T_j(G_i(f))(a) = L_{i+j}(f)(a)$, and $L_{i+j}\in \cal L$ as $i+j \leq s-1$. Hence, the pair $(\calT,\calG)$ \emph{list-composes} in terms of $\calL$. 
\cref{lem:general} is proved in \cref{sec:list-dec-loc}. We then use this theorem to demonstrate
that several families of linear operator codes are list-decodable up
to capacity in \cref{sec:examples}.

\section{List-decoding Polynomial Ideal Codes (Proof of {\cref{thm: decoding up to Johnson}})}\label{sec: poly ideal johnson bound}

In this section we prove \cref{thm: decoding up to Johnson}. The proof
proceeds in three steps. In the first step, we find a bivariate polynomial $Q(X, Y)$ such that for every polynomial $f  \in \F_{<k}[X]$, if $f\mod E_i = \vc_i$, then $Q(X, f(X)) = 0 \mod  E_i^r$. In the second step of the argument, we show that if $f$ is such that $\enc(f)$ is close enough to $\vc$, then $Q(X, f(X))$ must be the identically zero polynomial, and therefore, $(Y - f(X))$ is a factor of $Q(X, Y)$ in the ring $\F[X, Y]$. In the final step of the algorithm, we factor $Q(X, Y)$ to output all factors of the form $(Y-f(X))$, where $f$ has degree less than $k$ and $\enc(f)$ and $\vc$ have a large agreement. 

We now describe some of the details. 
\subsection{Interpolating a Polynomial of an Appropriate Form}
\begin{lemma}\label{lem: interpolation johnson}
Let $r \in \N$ be a parameter. Let $D$ be be an integer such that $(nsr(r+1)k)^{1/2} < D \leq (nsr(r+1)k)^{1/2} + 1$ and let $D' \geq D$ be an integer divisible by $s$. Then, there exist bivariate polynomials $Q(X, Y), \{B_{i}(X, Y) : i \in [n] \}$ and univariate polynomials $\{A_{i, j}(X) : i \in [n], j \in \{1, 2, \ldots, r\}\}$ such that the following conditions hold. 
\begin{itemize}
\item $Q$ is not identically zero. 
\item For every $i \in [n]$, 
\[
Q(X, Y) - (Y - \vc_i)^r\cdot B_i(X, Y) + \sum_{j = 1}^r E_i(X)^{j}(Y-\vc_i)^{r-j}\cdot A_{i, j}(X) = 0\, .
\]
\item The $(1, k-1)$-weighted degree of $Q$ is at most $D$. 
\item For each $i \in [n]$, the $(1,s)$-weighted degree of $B_i$ is at most $D'-rs$.
\item For each $i \in [n], j \in [r]\setminus \{0\}$, the degree of $A_{i, j}(X)$ is at most $D' - rs$. 
\end{itemize}
Moreover, these polynomials can be found deterministically in time $\poly(n, r, s)$. 
\end{lemma}
Before proceeding further, we remark that this slightly mysterious form of $Q$ in the \cref{lem: interpolation johnson} is to ensure that for any $f \in \F[X]$, and $i \in [n]$, if $f(X) \mod E_i(X) = \vc_i $, then $Q(X, f(X)) = 0 \mod E_i(X)^r$. We note this in the following claim. 
\begin{claim}\label{clm: props of Q}
Let $Q(X, Y)$, $\{B_i : i \in [n]\}$, $\{A_{i, j} : i \in [n], j \in \{1, 2, \ldots, r\}\}$ be polynomials satisfying the conditions in \cref{lem: interpolation johnson}. For any $f \in \F[X]$, and $i \in [n]$, if $f(X) \mod E_i(X) = \vc_i $, then \[
Q(X, f(X)) = 0 \mod E_i(X)^r \, .
\]
\end{claim}
\begin{proof}
The condition $f(X) \mod E_i(X) = \vc_i$ implies that $(f(X) - \vc_i) = 0 \mod E_i(X)$. Thus, for every  $j \in \{0,\ldots, r \}$, $E_i^j(X)\cdot (f(X) - \vc_i)^{r-j} = 0 \mod E_i(X)^r$. Therefore, in the expression,
\[
Q(X, f(X)) = (f(X) - \vc_i)^r\cdot B_i(X, Y) + \sum_{j = 1}^r E_i(X)^{j}(f(X)-\vc_i)^{r-j}\cdot A_{i, j}(X) \, ,
\]  
each of the summands is divisible by $E_i(X)^r$, and hence $Q(X, f(X)) = 0 \mod E_i(X)^r$. 
\end{proof}
We now move on to the proof of \cref{lem: interpolation johnson}. 
\begin{proof}[Proof of \cref{lem: interpolation johnson}]
The proof of the lemma is by a fairly standard argument of viewing the conditions in the second item of \cref{lem: interpolation johnson} as homogeneous linear constraints on the coefficients of the polynomials involved and observing that there are more variables than homogeneous linear constraints, and hence there is a non-zero solution which can be found algorithmically by standard linear algebra. One subtlety here is to note that any non-zero solution of this linear system leads to a non-zero polynomial $Q$ (required by the first item in \cref{lem: interpolation johnson}). This observation is crucial to ensure that $Q$ is non-zero, as apriori an arbitrary non-zero solution to this linear system could just mean that some of the coefficients of the other polynomials ($A_{i, j}$'s and $B_i$'s) are non-zero, but they somehow cancel each other out to ensure that $Q$ remains zero. We now argue that this cannot be the case. 

Let $i \in [n]$ be such that there is a solution to this linear system where $B_i, A_{i, 1}, \ldots, A_{i, r}$ are not all identically zero. If $Q$ is non-zero, then we are done. So, let us now assume that $Q$ is zero, and argue that this cannot be the case. If $B_i$ is non-zero, then, observe that $(Y-\vc_i)^rB_i(X, Y)$ contains a monomial with $Y$-degree at least $r$, and this cannot be cancelled by any monomial in $\sum_{j = 1}^r E_i(X)^{j}(f(X)-\vc_i)^{r-j}\cdot A_{i, j}(X)$ since the $\deg_Y$ for this polynomial is strictly less than $r$. Thus, $Q$ cannot be identically zero. Else, if $B_i$ is identically zero, then let $j \in \{1, 2, \ldots, r\}$ be the smallest index such that $A_{i, j}(X)$ is non-zero. Then, the summand $E_i(X)^j(Y - \vc_i)^{r-j}A_{i, j}(X)$ contains a non-zero monomial with $Y$-degree equal to $r -j$, which cannot be cancelled out by the rest of the summands. Therefore, $Q$ is non-zero. 

We now count the number of homogeneous linear constraints in the system. Since $s < k-1$, it follows that $Q$ must also have $(1, s)$-weighted degree at most $D$. Moreover, since $D \leq D'$, we have that for each $i$, the equation  
\[
Q(X, Y) - (Y - \vc_i)^r\cdot B_i(X, Y) + \sum_{j = 1}^r E_i(X)^{j}(Y-\vc_i)^{r-j}\cdot A_{i, j}(X) = 0 \, 
\]  
only involves monomials of $(1, s)$-weighted degree at most $D'$. Thus, from \cref{lem: wtd deg monomials}, each such linear constraint leads to at most $N(s, D')$  homogeneous constraints on the coefficients, where for natural numbers $a, b$, $N(a, b)$ denotes the number of bivariate monomials of $(1, a)$-weighted degree at most $b$.  Since there are $n$ such equations, the total number of homogeneous linear constraints is at most $nN(s, D')$. 

To get an upper bound on the number of variables in this homogeneous linear system, observe from the weighted degree conditions and \cref{lem: wtd deg monomials}  that the number of variables to this system contributed by $Q$ is at least $N(k-1, D)$, by each $B_i$ is at least $N(s, D'-rs)$ coefficients and each $A_{i,j}$ is at least $(D' - rs + 1)$ coefficients. Thus, the total number of variables is at least 
\[
N(k-1, D) + n\left(N(s, D'-rs) + (D' - rs +1)\right) \, .  
\]      
Thus, there exists a non-zero solution to this system of homogeneous linear equations if 
\[
N(k-1, D) + n\left(N(s, D'-rs) + r(D' - rs +1)\right) > nN(s, D') \, , 
\]
or, equivalently, 
\[
N(k-1, D) > n \left(N(s, D') - \left(N(s, D'-rs) + r(D' - rs +1)\right) \right)  \, .
\]
Now, from \cref{lem: wtd deg monomials}, we know that 
\[
N(k-1, D) \geq D^2/(2k) > nsr(r + 1)/2\, ,
\]
and 
\[
n \left(N(s, D') - \left(N(s, D'-rs) + r(D' - rs +1)\right) \right) = nsr(r + 1)/2 \, .
\]
This last inequality follows by invoking the second item of \cref{lem: wtd deg monomials} with $a = s, b = D', \eta = r$ (recall that $D'$ is divisible by $s$). So, we get that 
\[
n \left(N(s, D') - \left(N(s, D'-rs) + r(D' - rs +1)\right) \right) = nsr(r + 1)/2 \, .
\]
Thus, for our choice of parameters, we have 
\[
N(k-1, D) + n\left(N(s, D'-rs) + r(D' - rs +1)\right) > nN(s, D) \, , 
\]
and the system of equations must have a non-zero solution. 

We can find such a non-zero solution by solving the linear system, for instance, by Gaussian elimination over $\F$, which runs in time polynomial in the size of the system. This completes the proof of the lemma. 
\end{proof}

\subsection{Close Enough Codewords Satisfy the Equation}
We now prove the following lemma which is the second step for the proof of \cref{thm: decoding up to Johnson}. 

\begin{lemma}\label{lem: close enough johnson}
Let $D$ be as in \cref{lem: interpolation johnson} and let $Q(X, Y), \{B_{i} : i \in [n]\}, \{A_{i, j} : i \in [n], j \in \{1, 2, \ldots, r\}\}$ be  polynomials satisfying the conditions in \cref{lem: interpolation johnson}. And, let $f \in \F_{<k}[X]$ be such that $\enc(f)$ and $\vc$ agree on greater than $D/(rs)$ coordinates. Then, $Q(X, f(X))$ is identically zero. 
\end{lemma}
\begin{proof}
From \cref{clm: props of Q}, we know for any $i \in [n]$, $f(X) \mod E_i = \vc_i$ implies that $Q(X, f(X)) = 0 \mod E_i^r$. We also know from the statement of \cref{thm: decoding up to Johnson} that $E_0, E_1, \ldots, E_{n-1}$ are relatively prime. Therefore, if $S \subset [n]$ such that for all $i \in S$, $i \in [n]$, $f(X) \mod E_i = \vc_i$, then by the Chinese Remainder Theorem (see \cref{thm: CRT}), we have
\[
Q(X, f(X)) = 0 \mod \prod_{i \in S} E_i^r \, .
\]
We know that the degree of $Q(X, f(X))$ is at most the $(1, k-1)$-weighted degree of $Q$ which is at most $D$. Moreover, the degree of $\prod_{i \in S} E_i^r$ equals $|S|sr$, which is strictly larger than $D$ if $|S| > D/(rs)$. Thus, in this case, $Q(X, f(X)) = 0 \mod \prod_{i \in S} E_i^r$ implies that $Q(X, f(X))$ must be identically zero as a polynomial in $\F[X]$.
\end{proof}
\subsection{Reconstruction of All Close Enough Codewords}
Finally, from \cref{lem: close enough johnson}, we know that for any $f \in \F_{<k}[X]$ such that $\enc(f)$ and $\vc$ agree on at least $D/(rs)$ coordinates, $Q(X, f(X))$ must be identically zero. Thus, to recover all such $f$, we use any standard polynomial factorization algorithm (e.g. the algorithm due to Kaltofen \cite{Kaltofen1985}) to factor $Q(X, Y)$, and for every factor of the form $Y - f(X)$ such that $f(X)$ has degree less than $k$ and $\enc(f)$ and $\vc$ agree on greater than $D/(rs)$ coordinates, include $f$ in the output list. The list size is clearly bounded by the degree of $Q$, which is $\poly(n, r, s)$. 

Thus, we have an efficient algorithm which outputs all codewords which agree with the received word on greater than \[
D/(nrs) \leq 1/(nrs)\cdot ((nsr(r+1)k)^{1/2} + 1) \leq (1/(nrs) + (k/(sn))^{1/2}\cdot (1 + 1/r)^{1/2} )
\] fraction of coordinates. Choosing $r$ to be large enough, based on $\epsilon$, e.g. $r = \Theta(1/\epsilon)$, we get \cref{thm: decoding up to Johnson}. 

%\MKnote{a reference for factorization}

\section{List-decoding of Linear Operator
  Codes (Proof of {\cref{lem:general}})}\label{sec:list-dec-loc} 

In this section, we prove \cref{lem:general}. To this end, we follow the framework of Guruswami and Wang; the key observation being that the framework is general enough to be applicable to all families of codes with properties as stated in \cref{lem:general}, and not just Folded Reed-Solomon codes and Multiplicity codes, as shown by Guruswami and Wang. Before we proceed, we need some notation. 

For a natural number $n$, $[n]$ denotes the set $\{0, 1, \ldots, n-1\}$. Recall that the alphabet of the code is $\F^s$, and the block length is $|A| = n$. We denote the received word by $\vc \in {\F^s}^n$. For notational convenience, we  identify the set $[n]$ with the set $A$ via an arbitrary ordering of the elements of $A$. Thus, for every $a \in A$, we use $\vc_a \in \F^s$ to denote the $a$th coordinate of $\vc$. 

Recall that since $\calT$ is linearly-extendible, it follows that there exists a matrix $M_{\calT}$ such that for every polynomial $q(X) \in \F[X]$
 \[
 \calT( q(X) \cdot p(X) ) = q(M_{\calT}(X)) \cdot \calT(p(X)) \, .
 \] 
The proof of \cref{lem:general}, which follows the high level outline of the proof of Guruswami and Wang, follows from \cref{lem:interpolation,lem: close enough codewords satisfy the eqn,lem:reconstruction}. \cref{lem:interpolation} shows that we can interpolate a low degree polynomial $Q$, with appropriately nice structure and low enough degree, which \emph{explains} the received word $\vc$ in some sense. We then move on to observe in \cref{lem: close enough codewords satisfy the eqn} that any polynomial $f$ such that $\enc(f)$ is close enough to the received word $\vc$ in Hamming distance \emph{satisfies} an equation depending upon $Q$. Finally, in \cref{lem:reconstruction}, we \emph{solve} this equation, which is a  system of homogeneous linear equations on the coefficients of $f$ to recover all low degree polynomials $f$ such that $\enc(f)$ and $\vc$ are close enough. Because of the linear nature of constraints, all such solutions are contained in a low dimensional linear space. As we shall observe, each of these steps in the decoding procedure just involves doing some basic linear algebra over the underlying field, and hence the decoding can be done in polynomial time by a deterministic algorithm. 

We now proceed with the details of each of these steps. 
\subsection{Interpolating a Polynomial.}
\begin{lemma}\label{lem:interpolation}
There exists a non-zero polynomial $Q(X, U_0, U_1, \ldots, U_{w-1}) \in \F[X, \vu]$ of the form \[
Q(X, \vu) = \sum_{i = 0}^{w-1} Q_i(X)\cdot U_i \, ,
\] such that 
\begin{itemize}
\item For every $i \in [w]$, $\deg(Q_i)$ is at most $D = nr/w$. 
\item For every $a \in A$, 
\[
\left(\sum_{i \in [w]} Q_i(M_{\calT})(a) \cdot h_{G_i,a}\cdot \vc_a\right) = 0 \, ,
\]
\end{itemize}
where, (by a slight abuse of notation), we also use $h_{G_i, a}$ to denote the matrix associated with the linear transformation $h_{G_i, a}$.

Moreover, such a polynomial $Q$ can be constructed deterministically with at most $\poly(n)$ operations over the underlying field $\F$. 
\end{lemma}
\begin{proof}
The properties desired from $Q$ in the lemma can be viewed as a system of linear constraints on the coefficients of $Q$. More precisely, for every $a \in A$, the condition 
\[
\left(\sum_{i \in [w]} Q_i(M_{\calT})(a) \cdot h_{G_i,a}\cdot \vc_a\right) = 0
\]
imposes $r$ homogeneous linear constraints on the coefficients of $Q$. The existence of a non-zero polynomial $Q = \sum_{i \in [w]}Q_iU_i$ satisfying these constraints  now just follows from the fact that the number of homogeneous linear constraints is at most $nr$, whereas the number of variables is $(D + 1)w = (nr/w + 1)w > nr$. Thus, there is always a non-zero solution. 

Since the size of the linear system is polynomially bounded in $n$, a non-zero $Q$  satisfying the conditions can be found by solving the linear system, which can be done with $\poly(n)$ field operations using standard linear algebra algorithms. 
\end{proof}
\subsection{Close Enough Codewords Satisfy the Equation.}
We now argue that any polynomial $f \in \F[x]$ of degree at most $k-1$, whose encoding is close enough to the received word $\vc$ must satisfy an appropriate equation (depending upon $Q$). 
\begin{lemma}\label{lem: close enough codewords satisfy the eqn}
If $f \in \F[X]$ is a polynomial of degree less than $k$ such that for at least $n\cdot (1/w + k/(nr)) + 1$ points $a \in A$, $\enc(f)(a) = \vc_a$, then the polynomial $Q(X, G_0(f), G_1(f), \ldots, G_{w-1}(f)) \in \F[X]$ is identically zero. 
\end{lemma}
\begin{proof}
Let $R(X)$ be defined as 
\[
R(X) := Q(X, G_0(f), G_1(f), \ldots, G_{w-1}(f)) = \sum_{i \in [w]} Q_i(X) \cdot G_i(f) \, .
\]
Since $f$ is of degree at most $k-1$ and the operators $G_i$ do not increase the degree, $R$ is a polynomial of degree at most $D + k-1 \leq nr/w + k -1$. 

Let $a \in A$ be such that $\enc(f)(a) = \vc_{a}$, then, we will show that {$\calT(R)$ is zero at $a$}. Also, from the linearity of $\calT$, it follows that 
\[
\calT(R) = \calT\left(\sum_{i \in [w]} Q_i(x)\cdot G_i(f)\right) = \left(\sum_{i \in [w]}\calT\left( Q_i(X)\cdot G_i(f)\right)\right) \, .
\]
Using linear extendibility of $\calT$, we get 
\[
\calT(R) = \left(\sum_{i \in [w]} Q_i(M_{\calT}) \cdot \calT(G_i(f))\right) \, .
\] 
Now, since $(\calT, \calG)$ list composes in terms of $\calL$ at the set of evaluation points $A$, we know that for every $i \in [w]$, and $a \in A$, $\calT(G_i(f))(a) = h_{G_i,a}(\calL(f)(a))$. Therefore, 
\[
\calT(R)(a) = \left(\sum_{i \in [w]} Q_i(M_{\calT})(a) \cdot h_{G_i,a}(\calL(f)(a))\right) \, .
\]
Since $\enc(f)(a) = \calL(f)(a) = \vc_a$, we get that 
\[
\calT(R)(a) = \left(\sum_{i \in [w]} Q_i(M_{\calT})(a) \cdot h_{G_i,a}\cdot \vc_a\right) \, .
\]
Here, we abuse notation and also use $h_{G_i, a}$ to denote the matrix associated to the linear transformation given by $h_{G_i, a}$. Now, from the constraints on the polynomial $Q$ in \cref{lem:interpolation}, we know that the right-hand side of the above equation is zero for all $a \in A$. Thus, $\calT(R)(a)$ is zero, whenever $\enc(f)(a) = \vc_{a}$ for an $a \in A$.  

We now recall that since the operators $\calT$ give us a code with rate $(D+k)/(rn)  = 1/w + k/(rn)$ and distance $(1-1/w-k/(rn))$. Thus, if $\enc(f)$ and $\vc$ have agreed on greater than $(1/w+k/(rn))$ fraction of points in $A$, $R$ must be identically zero. 
\end{proof}

\subsection{Solving the Equation to Recover the Codewords}
We now show that we can solve equations of the form  
\[
Q(X, G_0(f), G_1(f), \ldots, G_{w-1}(f)) = 0 \, , 
\]
to recover a (small) list of all polynomials $f$ of degree at most $k-1$ which satisfy the above equation.
\begin{lemma}\label{lem:reconstruction}
The set of polynomials $f(X) \in \F[X]$ of degree at most $k-1$ such that the polynomial $Q_0(X)G_0(f) + Q_1(X)G_1(f) + \cdots + Q_{w-1}(X)\cdot G_{w-1}(f)$ is identically zero form a linear space of dimension at most $\ell$ over the underlying field $\F$. 

Moreover, there is a deterministic algorithm which runs in polynomial time and given $Q, \calG$ as input outputs a basis for this linear space. 
\end{lemma}
\begin{proof}
From the linearity of $\calG$, and the fact that $Q(X, \vu)$ is linear in the $u$ variables, it immediately follows that the set of polynomials $f$ of degree at most $k-1$ such that 
\[
Q_0(X)G_0(f) + Q_1(X)G_1(f) + \cdots + Q_{w-1}(X)\cdot G_{w-1}(f) \equiv 0
\]
form a linear space. Moreover, given the polynomial $Q$, and a description of $\calG$, we can set up this linear system in terms of the coefficients of $f$ and solve the system in time $\poly(n)$. So, all that remains for the proof of the lemma is to argue that the dimension of this solution space is not too large. For this, we will crucially rely on the property of $\calG$ that $\Diag(\calG)$ is the generator matrix of a code of distance $k-\ell$. We start with setting up some notation.

Let $d = \max_{j \in [w]} \deg(Q_j)$, and  $Q_j(X) = \sum_{i = 0}^d q_{j, i}X^i$. From the definition of $d$, it follows that the vector $\tilde{q} = (q_{0, d}, q_{1, d}, \ldots, q_{w-1, d})$ is \emph{not} the all zeros vector. For every $i, i' \in [k]$, let $g_{i}^j \in \F^{k}$ denote the $i$th row of the matrix $G_j$ (here we are interpreting $G_j:F^{<k}[X] \to F^{<k}[x]$ as a $kxk$ matrix) and let $g_{i, i'}^j$ denotes the $(i, i')$ element of $G_j$. We also note that since $\calG$ is a degree preserving set of linear operators, each of these matrices $G_j$ are upper triangular (here we interpret $G_j$ acting on vectors whose $i^{\text{th}}$ coordinate corresponds to the coefficient of $X^i$ and so on). Our goal is to find the set of all vectors $\coeff(f) = (f_{k-1}, f_{k-2}, \ldots, f_{0})$ where $f = \sum_{j = 0}^{k-1} f_jx^j$ satisfies the equation
\[
Q_0(X)G_0(f) + Q_1(X)G_1(f) + \cdots + Q_{w-1}(X)\cdot G_{w-1}(f) \equiv 0 \, .
\]
We note that this is equivalent to saying that the coefficient of every monomial in $X$ on the left hand side is zero. Moreover, the $Q_0(X)G_0(f) + Q_1(X)G_1(f) + \cdots + Q_{w-1}(X)\cdot G_{w-1}(f)$ is a polynomial of degree $d + k-1$. We now chase down some of these coefficients  in decreasing order of their degree, as summarised in the following simple claim. %For instance, we observe that the coefficient of $Y^{d + k-1}$ equals $\sum_{j \in [w]} q_{j, d}\cdot \langle q_{j, k}, \coeff(f)\rangle$ and the coefficient of $Y^{d + k-2}$ equals $\sum_{j \in [w]} q_{j, d}\cdot \langle q_{j, k-1}, \coeff(f)\rangle + \sum_{j \in [w]} q_{j, d-1}\cdot \langle q_{j, k}, \coeff(f)\rangle$. More generally, we have the following simple calim.
\begin{claim}\label{clm:chasing coefficients}
For each $i < k$, the coefficient of $X^{d + k-1-i}$ in $Q_0(X)G_0(f) + Q_1(X)G_1(f) + \cdots + Q_{w-1}(X)\cdot G_{w-1}(f)$ equals 
\[
\sum_{j \in [w]} \left( \sum_{i'=0}^i  q_{j, d-i'}\cdot \langle g_{k-1-(i-i')}^j, \coeff(f)\rangle \right) \, .
\]
\end{claim}
From the degree preserving property of $\calG$, we also know that the coefficient of $X^{d+k-1-i}$ in the above claim only depends on $f_{k-1-i}, f_{k-i}, \ldots f_{k-1}$. In particular, if we set up a linear system where the $i$th constraint equates the coefficient of $X^{d+k-1-i}$ obtained in \cref{clm:chasing coefficients} to zero, then resulting linear system in $(f_{k-1}, f_{k-2}, \ldots, f_1, f_0)$ is lower triangular, and the diagonal elements of the  matrix of linear constraints which equals the coefficient of $f_{k-1-i}$ in the expression $\sum_{j \in [w]} \left( \sum_{i'=0}^i  q_{j, d-i'}\cdot \langle g_{k-1-(i-i')}^j, \coeff(f)\rangle \right)$ is precisely
\[
\sum_{j \in [w]} q_{j, d}g_{k-1-i, k-1-i}^j \, .
\]
We can view $\sum_{j \in [w]} q_{j, d}g_{k-1-i, k-1-i}^j$ as an inner product of the (non-zero) vector $\tilde{q} = (q_{0,d}, \ldots, q_{w-1,d})$ with the vector $v_i = (g_{k-1-i, k-1-i}^0, \ldots, g_{k-1-i, k-1-i}^{w-1})$. Now, we know that $q$ is a non-zero vector. So, if we can ensure that at most $\ell$ of the coordinates of the vector $(\langle \tilde{q},v_i \rangle : i \in [k])$ are zero, we would have the desired bound of $\ell$ on the dimension of the solution space. 

To this end, consider the $k \times w$ matrix $W$, whose $i$th row is the vector $v_i$. From the definition of $v_i$, we can observe that the $j$th column of this matrix are precisely the main diagonal of the matrix $G_j$. From the last item in the hypothesis of \cref{lem:general}, we know that this matrix $W$ is code of distance $k-\ell$. Thus,  the vector $W\cdot \tilde{q}$ can be zero on at most $\ell$ of its coordinates. This gives us an $\ell$ dimensional linear space containing all the solutions $f$ of this equation, and therefore a bound of $|\F|^{\ell}$ on the  size of this solution space.  
\end{proof}
\begin{remark}
 {Notice that in \cref{lem:reconstruction} we recovered the coefficients of the polynomial $f$ in decreasing order of their degree. The advantage to recovering the coefficients in this order is as follows. The degree preserving nature of the $G$'s ensures that the coefficient $f_i$ doesn't play a role in the coefficient of $X^{d'}$ in $Q_0(X)G_0(f) + Q_1(X)G_1(f) + \cdots + Q_{w-1}(X)\cdot G_{w-1}(f)$ when $d'$ is larger that $d-i$. This leads to a triangular system of equations whose rank can be easily inferred from the diagonal elements.}
\end{remark}

\section{Example of Codes Achieving List-Decoding Capacity }\label{sec:examples}
	
	\SBnote{Somewhere in the write-up we need to add that while additive FRS can be got via Swastik's trick, affine FRS still slip the trick }

        In this section we will use \cref{lem:general} to (re)prove
        the list-decoding capacity of the Folded Reed-Solomon codes,
        multiplicity codes and additive Folded Reed-Solomon codes. We
        then introduce a common generalization of all these codes,
        which we refer to as affine Folded Reed-Solomon codes and
        prove the list-decoding up to capacity of these codes.

        We recall that Guruswami and Rudra \cite{GuruswamiR2008} proved the list-decoding capacity of FRS codes, first introduced by Krachkovsky \cite{Krachkovsky2003}
        while Kopparty \cite{Kopparty2015} proved the list-decoding
        capacity of multiplicity codes. Guruswami and Wang
        \cite{GuruswamiW2013} then gave an alternative and simpler
        linear-algebraic framework to prove the list-decoding capacity
        of both FRS and multiplicity codes. The list-decoding
        capacity of additive FRS codes is proved using the more
        involved algorithm of Guruswami \& Rudra \cite{GuruswamiR2008}
        and an observation of Kopparty \cite{Kopparty2015} (see paragraph on Additive
        Folding and Footnote $4$ in \cite[Section
        III]{Guruswami2011}). More recently (subsequent to the
        conference version of this paper), Gopi and Guruswami
        \cite{GopiG2022} used skew polynomials to construct improved
        maximally recoverable local reconstruction codes (MR LRCs). It can be shown that the
        machinery of skew polynomials can be used to give yet another
        proof of list-decodability of FRS, multiplicity and additive
        FRS codes (see \cite[Appendix C]{GopiG2022}). We remark that it is apriori unclear how to prove
        list-decodability of affine FRS codes using skew polynomials
        (or via the previous frameworks of \cite{GuruswamiR2008,
          Kopparty2015, GuruswamiW2013}).
        
	\subsection{Folded Reed-Solomon ($FRS$) Codes} 
	
	Fix integers $k,n,q$ with $n\leq q$. Fix $\gamma\in \F_q^{*}$
        of multiplicative order at least $s$. The message space of the $FRS^{\gamma}_s[k,n]$ code with folding parameter $s$ is polynomials of degree at most $k-1$ over $\F[X]$, i.e., $\F_{<k}[X]$ where $\F=\F_q$.  Then, $FRS$ codes are linearly-extendible linear operator codes $LELO_{\calL,A}$ where:
	\begin{itemize}
	    \item $\mathcal{L}=(L_0,\ldots,L_{s-1})$ with $L_1(f(X))= f(\gamma X)$  for $f(X)\in \F_q[X]$ and $L_i=L_1^i$ for $i \in \set{0,1,\ldots,s-1}$.  
	    \item For the above family of operators $M(X)$ is given by $M(X)_{ij}=\gamma^{i}X\cdot \mathbb{I}[i=j]$ for $i,j\in [s]$.
	    \item The set of evaluation points is $A=\set{a_0,\ldots,a_{n-1}}$ where for any two distinct $i$ and $j$ the sets $\set{a_i,a_i\gamma,\ldots,a_i\gamma^{s-1}}$ and $\set{a_j,a_j\gamma,\ldots,a_j\gamma^{s-1}}$ are disjoint.
	\end{itemize}
	
	\begin{remark}\;
	  \begin{enumerate}
	  \item Recall that the bivariate polynomial $E(X,Y)$ corresponding to the polynomial ideal code representation is $E(X,Y)=\prod_{i=0}^{s-1}(X-\gamma^iY)$.
      \item For the choice of $A$ as above, the rate of the code is $\frac{k}{sn}$ and its distance is $1-\frac{k-1}{sn}$ as the polynomials $E_i=E(X,a_i)$ are pairwise co-prime.
	  \end{enumerate}
	\end{remark}
	
	\begin{theorem}[\cite{GuruswamiW2013}]
	 Let $\gamma\in \F_q^{*}$ be an element of order at least $k$. Further, let $A=\set{a_0,\ldots,a_{n-1}}$ be a set of evaluation points where for any two distinct $i$ and $j$ the sets $\set{a_i,a_i\gamma,\ldots,a_i\gamma^{s-1}}$ and $\set{a_j,a_j\gamma,\ldots,a_j\gamma^{s-1}}$ are disjoint.
	 For every $\epsilon>0$ there exists $s$ large enough ($s\geq
         \Omega(1/\epsilon^2)$) such that $FRS_s^{\gamma}[k,n]$ at the
         set of evaluation points $A$ can be efficiently list-decoded
         up to distance $1-\frac{k}{sn}-\epsilon$.
	\end{theorem} 
	\begin{proof}
	We will prove this by applying \cref{lem:general}. Set $\mathcal{G}=(L_0,\ldots,L_{w-1})$ for some integer $w<s$ to be set later and $\calT=(T_0,\ldots,T_{r-1})$ with $r=s-w+1$ and $T_i=L_i$. 
	
	\cref{lem:general}-\cref{itm:general_lem_lin_ext}: Clearly, $(\calT,A)$ forms a linearly-extendible linear operator code $LELO_{k+nr/w}^A(\calT)$ which is $FRS_r^{\gamma}[k+nr/w,n]$ at the set of evaluation points $A$.
	
	\cref{lem:general}-\cref{itm:general_lem_list_comp}: 
	For all $G_i \in \calG$, $T_j\in \calT$ and $a \in A$, we have that for every polynomial $f \in \F[X]$: $T_j(G_i(f))(a) = L_{i+j}(f)(a)$. Notice that $L_{i+j}\in \cal L$ as $i+j \leq s-1$.
	
	\cref{lem:general}-\cref{itm:general_lem_deg_pre}: $G_i(x^j)=\gamma^{ij}X^j$, and hence $\calG$ is degree preserving.	
	
	\cref{lem:general}-\cref{itm:general_lemma_G_dist}:  
	The matrix $\Diag(\mathcal{G})$ is given by $\Diag(\mathcal{G})_{ij}=\gamma^{ij}$ for $i\in [w]$ and $j \in [k]$. Hence, as long as $\gamma$ has order at least $k$ this is the generator matrix of $RS[w-1,k]$ and hence its distance is $k-w+1$. 

	Thus $FRS_s^\gamma[k,n]$ can be efficiently list-decoded up to
        distance $1-\frac{k-1}{rn}-\frac{1}{w}$ with list size $q^{w-1}$.
	By choosing a large enough $w$ and $s$ we can ensure that $1-\frac{k-1}{rn}-\frac{1}{w}>1-\frac{k}{sn}-\epsilon$.
	\end{proof}

\subsection{Multiplicity ($MULT$) Codes}

	Fix integers $k,n,q$ with $n\leq q$. The message space of the $MULT_s[k,n]$ code of order $s$ is polynomials of degree at most $k-1$ over $\F[X]$, i.e., $\F_{<k}[X]$ where $\F=\F_q$. Then, $MULT_s[k,n]$ codes are linearly-extendible linear operator codes $LELO_{\calL,A}$ where:
	\begin{itemize}
	    \item $\mathcal{L}=(L_0,\ldots,L_{s-1})$ with $L_1(f(X))= \frac{\partial f(X)}{\partial X}$  for $f(X)\in \F_q[X]$ and $L_i=L_1^i$ for $i \in \set{0,1,\ldots,s-1}$.  
	    \item For the above family of operators $M(X)$ is given by $M(X)_{ij}=X\cdot\mathbb{I}[i=j] + i\cdot \mathbb{I}[i-1=j]$ for $i,j\in [s]$.
	    \item The set of evaluation points is $A=\set{a_0,\ldots,a_{n-1}}$ where $a_i$s are all distinct.
	\end{itemize}
	
	\begin{remark}\;
	  \begin{enumerate}
	      \item  Recall that the bivariate polynomial $E(X,Y)$ corresponding to the polynomial ideal code representation is $E(X,Y)=(X-Y)^{s}$.
	      \item For the choice of $A$ as above, $MULT_s[k,n]$ is a code with rate $\frac{k}{sn}$ and distance $1-\frac{k-1}{sn}$ as the polynomials $E_i=E(X,a_i)$ are pairwise co-prime.
	  \end{enumerate}
	\end{remark}
	
	\begin{theorem}[\cite{GuruswamiW2013}]
	\label{thm:MULT}
	Let the characteristic of $\F_q$ be at least $\max(s,k)$. Further, let the set of evaluation points be $A=\set{a_0,\ldots,a_{n-1}}$ where $a_i$s are all distinct.
	 Then, for every $\epsilon>0$ there exists $s$ large enough
         ($s\geq \Omega(1/\epsilon^2)$) such that $MULT_s[k,n]$ can be
         efficiently list-decoded up to distance $1-\frac{k}{sn}-\epsilon$. 
	\end{theorem} 
	\begin{proof}
	We will again appeal to \cref{lem:general}. Set $\calG=(G_0,\ldots,G_{w-1})$ where $G_i=\frac{X^i}{i!}\cdot L_i$ for $i\in \set{0,1,\ldots,w-1}$ for some integer $w<s$ to be set later and $\calT=(T_0,\ldots,T_{r-1})$ with $r=s-w+1$ and $T_i=L_i$. 
	
	\cref{lem:general}-\cref{itm:general_lem_lin_ext}: Clearly, $(\calT,A)$ forms a linearly-extendible linear operator code $LELO_{k+nr/w}^A(\calT)$ which is $MULT_r[k+nr/w,n]$ of order $r$ at the set of evaluation points $A$.
	
	\cref{lem:general}-\cref{itm:general_lem_list_comp}: 
	For all $G_i \in \calG$, $T_j\in \calT$ and $a \in A$, we have that for every polynomial $f \in \F[X]$: \[T_j(G_i(f))(a) = (\sum_{b=0}^{j}\binom{j}{b}\binom{i}{b}\cdot (b!/i!)\cdot X^{i-b}L_{i+b}(f))(a).\]
	Notice that the above expression only involves $L_i$s where $i<s$.
	
	\cref{lem:general}-\cref{itm:general_lem_deg_pre}: $G_i(X^j)=\binom{j}{i} \cdot X^j$, and hence $\calG$ is degree preserving.	
	
	\cref{lem:general}-\cref{itm:general_lemma_G_dist}:  
    The matrix $\Diag(\mathcal{G})$ is given by $\Diag(\mathcal{G})_{ij}=\binom{j}{i}$ for $i\in [w]$ and $j\in [k]$. This matrix can be transformed via elementary row operations to a $RS[w,k]$ generator matrix with points of evaluations as $0,1,\ldots,k-1$; thus, as long as the characteristic of $\F_q$ is at least $k$ we have that the distance of $\Diag(\mathcal{G})$ is $k-w+1$.
    
	Thus $MULT_s[k,n]$ can be efficiently list-decoded up to
        distance $1-\frac{k-1}{rn}-\frac{1}{w}$ with list size $q^{w-1}$.
	By choosing a large enough $w$ and $s$ we can ensure that $1-\frac{k-1}{rn}-\frac{1}{w}>1-\frac{k}{sn}-\epsilon$.
\end{proof}

\subsection{Additive Folded Reed-Solomon ($\AFRS$) Codes}
    	Fix integers $k,n,q$ with $n\leq q$. Let $\beta\in \F_q$ be a
        non-zero element and characteristic of $\F_q$ is at least $s$. The message space of the $\AFRS_s^{\beta}[k,n]$ code with folding parameter $s$ is polynomials of degree at most $k-1$ over $\F[X]$, i.e., $\F_{<k}[X]$ where $\F=\F_q$. Then, $\AFRS_s^\beta[k,n]$ codes are linearly-extendible linear operator codes $LELO_{\calL,A}$ where:
	\begin{itemize}
	    \item $\mathcal{L}=(L_0,\ldots,L_{s-1})$ with $L_1(f(X))= f(X+\beta)$  for $f(X)\in \F_q[X]$ and $L_i=L_1^i$ for $i \in \set{0,1,\ldots,s-1}$.  
	    \item For the above family of operators $M(X)$ is given by $M(X)_{ij}=(X+i\beta)\cdot \mathbb{I}[i=j]$ for $i,j\in [s]$.
	    \item The set of evaluation points is $A=\set{a_0,\ldots,a_{n-1}}$ where $a_i-a_j\notin \set{0,\beta,2\beta,\ldots,(s-1)\beta}$ for distinct $i$ and $j$.
	\end{itemize}
	
	\begin{remark}\;
	  \begin{enumerate}
	  \item Recall that the bivariate polynomial $E(X,Y)$ corresponding to the polynomial ideal code representation is $E(X,Y)=\prod_{i=0}^{s-1}(X-Y-i\beta)$.
      \item For the choice of $A$ as above, $\AFRS_s^{\beta}[k,n]$ is a code with rate $\frac{k}{sn}$ and distance $1-\frac{k-1}{sn}$ as the polynomials $E_i=E(X,a_i)$ are pairwise co-prime.
	  \end{enumerate}
	\end{remark}
	\begin{theorem}
	\label{thm:AFRS}
	 Let the characteristic of $\F_q$ be at least $\max(s,k)$ and $\beta \in \F_q$ be a non-zero element.
	 Further, let the set of evaluation points  $A=\set{a_0,\ldots,a_{n-1}}$ be such that $a_i-a_j\notin \set{0,\beta,2\beta,\ldots,(s-1)\beta}$ for distinct $i$ and $j$.
	 Then, for every $\epsilon>0$ there exists $s$ large enough
         ($s\geq \Omega(1/\epsilon^2)$) such that
         $\AFRS_s^{\beta}[k,n]$ over the set of evaluation points $A$
         can be efficiently list-decoded up to distance $1-\frac{k}{sn}-\epsilon$.
	\end{theorem} 
	\begin{proof}
	We will again appeal to \cref{lem:general}. To define $\mathcal{G}=(G_0,\ldots,G_{w-1})$ for some integer $w<s$, we need the following definitions. Let $B\in \F_q^{w\times w}$ be a matrix where $B_{ij}=(j)^{i}$ for $i,j\in [w]$, i.e, the transpose of the Vandermonde matrix at the points $\set{0,1,\ldots,w-1}$: these points are distinct since the characteristic of the field is at least $k$. Further, let $\mathbf{b}_i \in \F_q^w$ be a vector such that $B \mathbf{b}_i = e_i$ for $i\in [w]$ where $e_i$s are the standard basis vectors: $\mathbf{b}_i$s exist because $B$ is full rank.
	Now, define $G_{i}=X^{i}\cdot \sum_{c=0}^{w-1} \mathbf{b}_i(c)L_{c}$ for $i\in [w]$. Set $\calT=(T_0,\ldots,T_{r-1})$ with $r=s-w+1$ and $T_i=L_i$. 
	
	\cref{lem:general}-\cref{itm:general_lem_lin_ext}: Clearly, $(\calT,A)$ forms a linearly-extendible linear operator code $LELO_{k+nr/w}^A(\calT)$ which is $\AFRS_r^\beta[k+nr/w,n]$ with folding parameter $r$ at the set of evaluation points $A$.
	
	\cref{lem:general}-\cref{itm:general_lem_list_comp}: 
	For all $G_{i} \in \calG$, $T_j\in \calT$ and $a \in A$, we have that for every polynomial $f \in \F[X]$: 
	\begin{align*}
	    	T_j(G_{i}(f))(a) = T_j\left(X^{i}\cdot\sum_{c=0}^{w-1}  \mathbf{b}_{i}(c)L_{c}\right)(a)\\
	= \left((X+j\beta)^{i}\cdot\sum_{c=0}^{w-1} \mathbf{b}_{i}(c)L_{c+j}\right)(a).
	\end{align*}
Notice that the above expression only involves $L_i$s where $i<s$. 
	\cref{lem:general}-\cref{itm:general_lem_deg_pre}:
	\begin{align*}
	    G_{i}(X^j)&=X^{i}\cdot \sum_{c=0}^{w-1} \mathbf{b}_i(c)L_{c}(X^j)\\
	    &=X^{i}\cdot \sum_{c=0}^{w-1}\mathbf{b}_i(c) (X+c\beta)^j\\
	    &=X^{i}\cdot \sum_{c=0}^{w-1}\mathbf{b}_i(c)\sum_{h\leq j} \binom{j}{h}X^{h}\cdot (c\beta)^{j-h}\\
	    &=X^{i}\cdot \left(\binom{j}{i}\beta^{i}X^{j-i}+\sum_{h\leq j-w} \alpha_h X^{h}\right)\\
	    \intertext{(this is because $B \mathbf{b}_i = e_i$ which weans that for $h>j-w$ we have $\sum_{c=0}^{w-1} \mathbf{b}_i(c)\cdot (c)^{j-h} = \mathbb{I}[j-h=i]$; $\alpha_h$ are field constants)}
	    &=\binom{j}{i}\beta^{i-1}X^{j}+\ldots,
	\end{align*}
	and hence $\calG$ is degree preserving.	
	
	\cref{lem:general}-\cref{itm:general_lemma_G_dist}:  
	By the above, the matrix $\Diag(\mathcal{G})$ is given by $\Diag(\mathcal{G})_{ij}=\binom{j}{i}\beta^{i}$ for $i\in [w]$ and $j \in [k]$. Up to scaling this is the same code as $\Diag(\calG)$ in \cref{thm:MULT}: and hence, if the characteristic of the field is at least $k$ then its distance is $k-w+1$. 

	Thus $\AFRS_s^{\beta}[k,n]$ can be efficiently list-decoded up
        to distance $1-\frac{k-1}{rn}-\frac{1}{w}$ with list size $q^{w-1}$.
	By choosing a large enough $w$ and $s$ we can ensure that $1-\frac{k-1}{rn}-\frac{1}{w}>1-\frac{k}{sn}-\epsilon$.
	\end{proof}

\subsection{Affine Folded Reed-Solomon ($\text{Affine-FRS}$) Codes}
    We first recall the defintion of $\text{Affine-FRS}$ codes.
    	Fix integers $k,n,q$ with $n\leq q$. Let $\alpha \in \F_q^*$
        and $\beta\in \F_q$ such that the multiplicative order of $\alpha$ is $u$. 
    	Further, define $\ell(X)=\alpha X+\beta$ and 
    	\[
    	\ell^{(i)}(X)= \underbrace{\ell(\ell\ldots \ell(X))}_{i \text{
    times}} = \alpha^iX +\beta\cdot\sum_{j=0}^{i-1}\alpha^{j} = \alpha_iX + \beta_i.
    \]
    	In fact, if $\alpha\neq 1$, i.e, $u>1$ then,
        $\ell^{(u)}(X)=\ell^{(0)}(X)$. Let $\ord(\ell)$ denote the smallest positive integer $t$
    such that $\ell^{(t)}(z) = z$.
    	The message space of the $\text{Affine-FRS}_s^{\alpha,\beta}[k,n]$ code with folding parameter $s$ is polynomials of degree at most $k-1$ over $\F[X]$, i.e., $\F_{<k}[X]$ where $\F=\F_q$. Let the set of evaluation points be $A=\set{a_0,\ldots,a_{n-1}}$ such that for distinct $i,j$ the sets $\set{\ell^{(0)}(a_i),\ldots,\ell^{(s-1)}(a_i)}$ and $\set{\ell^{(0)}(a_j),\ldots,\ell^{(s-1)}(a_j)}$ are disjoint.
    	Then, $\text{Affine-FRS}_s^{\alpha,\beta}[k,n]$ codes are polynomial ideal codes where:
    	\begin{itemize}
    	\item The bivariate polynomial $E(X,Y)$ corresponding to the polynomial ideal code representation is $E(X,Y)=\prod_{i=0}^{s-1}(X-\alpha_iY -\beta_i)$.
        \item For the choice of $A$ as above, $\AfFRS_s^{\alpha,\beta}[k,n]$ is a code with rate $\frac{k}{sn}$ and distance $1-\frac{k-1}{sn}$ as the polynomials $E_i=E(X,a_i)$ are pairwise co-prime.
    	\end{itemize}

%     	linearly-extendible linear operator codes $LELO_{\calL,A}$ where:
% 	\begin{itemize}
% 	    \item $\mathcal{L}=(L_0,\ldots,L_{s-1})$ with $L_1(f(X))= f(\alpha Y+\beta)$  for $f(X)\in \F_q[X]$ and $L_i=L_1^i$ for $i \in \set{0,1,\ldots,s-1}$.  
% 	    \item For the above family of operators $M(Y)$ is given by $M(Y)_{ij}=(\alpha_iY+\beta_i)\cdot \mathbb{I}[i=j]$ for $i,j\in [s]$.
% 	    \item The set of evaluation points is $A=\set{a_0,\ldots,a_{n-1}}$ such that for distinct $i,j$ the sets $\set{\ell^{(0)}(a_i),\ldots,\ell^{(s-1)}(a_i)}$ and $\set{\ell^{(0)}(a_j),\ldots,\ell^{(s-1)}(a_j)}$ are disjoint.
% 	\end{itemize}
	
% 	\begin{remark}~\\
% 	  \begin{enumerate}
% 	  \item Recall that the bivariate polynomial $E(X,Y)$ corresponding to the polynomial ideal code representation is $E(X,Y)=\prod_{i=0}^{s-1}(X-\alpha_iY -\beta_i)$.
%       \item For the choice of $A$ as above, $AfFRS_s^{\alpha,\beta}[k,n]$ is a code with rate $\frac{k}{sn}$ and distance $1-\frac{k-1}{sn}$ as the polynomials $E_i=E(X,a_i)$ are pairwise co-prime.
% 	  \end{enumerate}
% 	\end{remark}

We will now recall the description of $\AfFRS$ codes in terms of linear operators which will be helpful while list-decoding. Define $D_1:\F[X]\to \F[X]$ as $D_1(f(X))=\frac{\partial f(X)}{\partial X}$ and $S_1:\F[X]\to \F[X]$ as $S_1(f(X))=f(\ell(X))$. Further, for $i\geq 0$ let $D_i=D_1^i$ and $S_i=S_1^i$. Recall, that the order of $\alpha$ is $u$. For any integer $r\in[s]$ let $r=r_1u+r_0$, with $r_0<u$, be the unique representation of $r$. Then, define $L_{r}:\F[X]\to \F[X]$ as $L_{r}(f(X))=S_{r_0}(D_{r_1}f(X))$. Set $\calL=(L_0,\ldots,L_{s-1})$. Clearly, $\calL$ is a family of linear operators. Further, $L_{r}(Xf)=S_{r_0}(D_{r_1}Xf)=S_{r_0}(r_1\cdot D_{r_1-1}f+X\cdot D_{r_1}f)=r_1\cdot L_{r-u}f+S_{r_0}(X)\cdot L_rf$: hence, $\calL$ is a set of linearly-extendible linear operators.

\begin{observation}\label{obs:affrs_lin_op_rep}
  If $u>1$ then at an evaluation point $a\in F_q$ the following pieces of information are the same:
  \begin{itemize}
      \item $f(X) \mod \prod_{i=0}^{s-1}(X-\alpha_ia -\beta_i) $
      \item $\calL(f)(a)$.
  \end{itemize}
\end{observation}
Hence, if $u>1$, then, $\AfFRS_s^{\alpha,\beta}[k,n]$ at the points of evaluation $A$ is $LELO_{\calL,A}$.

\begin{theorem}\label{thm:afFRS} For every $\epsilon >0$, there exists a large enough
  $s$ such that the follow holds. Let $\F_q$ be a field, $k$ a
  parameter and $\ell(X) = \alpha\cdot X + \beta$ such that $\alpha\in \F_q^{*}$ and $\beta \in \F_q $. Furthermore, let the
  evaluation points $A=\set{a_0,\ldots,a_{n-1}}$ be such that for
  distinct $i,j$ the sets $\set{\ell^{(0)}(a_i),\ldots,\ell^{(s-1)}(a_i)}$
  and $\set{\ell^{(0)}(a_j),\ldots,\ell^{(s-1)}(a_j)}$ are disjoint. Then, if either:
  \begin{itemize}
      \item $\ord(\ell)\geq k$ or
      \item $\charac(\F_q)>k$ and $\beta\neq 0$
  \end{itemize}
  holds, $\AfFRS_s^{\alpha,\beta}[k,n]$ over the set of evaluation points $A$
  can be efficiently list-decoded up to distance $1-\frac{k}{sn}-\epsilon$.
	\end{theorem} 
	\begin{proof}
	 We will again appeal to \cref{lem:general}. Let $u$ be the multiplicative order of $\alpha$.  Let $v=\floor{s/u}$.

        % \paragraph{Case $u=1$ and $\beta=0$:} This is the same case as for $MULT$ codes. Thus, by \cref{thm:MULT} we are done.
        
        \paragraph{Case $\ord(\ell)\geq k$:} This means that $u\geq k$. This is similar to decoding $FRS$ codes. We skip the details.
     
     \textit{Henceforth, we assume that $\charac(\F_q)\geq k$ and $\beta \neq 0$.}
        \paragraph{Case $u=1$:} This is the same case as for $\AFRS$ codes. Thus, by \cref{thm:AFRS} we are done.
	
	\paragraph{Case $u>1$ and $v\geq \sqrt{s}$:} (This case is similar to $MULT_v[k,n]$.)
	
	Define $\mathcal{G}=(G_0,\ldots,G_{w-1})$ for some integer $w<s$, as $G_i(f)=(X^i/i!)\cdot D_if$. Let $r=(v-w)u$ and set $\calT=\set{L_0,L_1,\ldots,L_{r-1}}$.
	
	\cref{lem:general}-\cref{itm:general_lem_lin_ext}: Clearly, $(\calT,A)$ forms a linearly-extendible linear operator code $LELO_{k+nr/w}^A(\calT)$ which is $\AfFRS_r^{\alpha,\beta}[k+nr/w,n]$ at the set of evaluation points $A$.
	
	\cref{lem:general}-\cref{itm:general_lem_list_comp}: 
	For all $G_{i} \in \calG$, $T_j\in \calT$ and $a\in A$ we have that for every polynomial $f \in \F[X]$: 
	\begin{align*}
	    	T_{j}(G_{i}(f))(a) &= \left(S_{j_0}D_{j_1}(\frac{X^i}{i!}\cdot D_i(f))\right)(a)\\
	    	&=\left(S_{j_0}\sum_{b=0}^{j_1}\binom{j_1}{b}\binom{i}{b}\cdot(b!/i!)\cdot X^{i-b}D_{i+b}(f)\right)(a)\\
	    	&=\left(\sum_{b=0}^{j_1}\binom{j_1}{b}\binom{i}{b}\cdot(b!/i!)\cdot (S_{j_0}X^{i-b})\cdot L_{j_0+(i+b)u}(f)\right)(a).
	\end{align*}
Notice that the above expression only involves $L_i$s where $i<s$. 

	\cref{lem:general}-\cref{itm:general_lem_deg_pre,itm:general_lemma_G_dist}: are identical to the corresponding items in \cref{thm:MULT}.
	
	Thus $\AfFRS_s^{\beta}[k,n]$ can be efficiently list-decoded up
        to distance $1-\frac{k-1}{rn}-\frac{1}{w}$ with list size $q^{w-1}$.
	By choosing a large enough $w$ and $s$ we can ensure that $1-\frac{k-1}{rn}-\frac{1}{w}>1-\frac{k}{sn}-\epsilon$.

	\paragraph{Case $u>\sqrt s$:}
	(This case is similar to $\AFRS_u^\beta[k,n]$.)
	 As in \cref{thm:AFRS}, to define $\mathcal{G}=(G_0,\ldots,G_{w-1})$ for some integer $w<u$, we need the following definitions. 
	Let $B\in \F_q^{w\times w}$ be a matrix where $B_{ij}=(\beta(\alpha^{j}-1)/(\alpha^j))^{i}$ for $i,j\in [w]$, i.e, the transpose of the Vandermonde matrix at the points $\set{\beta(\alpha^{j}-1)/(\alpha^j)\mid j\in [w]}$: these points are distinct since the order of $u$ is at least $w$. Further, let $\mathbf{b}_i \in \F_q^w$ be a vector such that $B \mathbf{b}_i = e_i$ for $i\in [w]$ where $e_i$s are the standard basis vectors: $\mathbf{b}_i$s exist because $B$ is full rank.
	
	Define $\calG=(G_0,\ldots,G_{w-1})$ for some integer $w<s$, as $G_i=X^i\cdot \sum_{c=0}^{w-1}b_i(c)S_c$. Let $r=s-w+1$ and
	set $\calT=\set{L_0,\ldots,L_{r-1}}$.

	\cref{lem:general}-\cref{itm:general_lem_lin_ext}: Clearly, $(\calT,A)$ forms a linearly-extendible linear operator code $LELO_{k+nr/w}^A(\calT)$ which is $\AfFRS_r^{\alpha,\beta}[k+nr/w,n]$ at the set of evaluation points $A$.
	
	\cref{lem:general}-\cref{itm:general_lem_list_comp}: 
	For all $G_{i} \in \calG$, $T_j\in \calT$ and $a\in A$ we have that for every polynomial $f \in \F[X]$: 
	\begin{align*}
	    	T_{j}(G_{i}(f))(a) &= \left(S_{j_0}D_{j_1}\left(X^i\cdot \sum_{c=0}^{w-1}b_i(c)S_cf\right)\right)(a)\\
	    	&=\left(S_{j_0}\sum_{b=0}^{j_1}\binom{j_1}{b}\binom{i}{b}\cdot(b!)\cdot X^{i-b}D_{b}\left(\sum_{c=0}^{w-1}b_i(c)S_cf\right)\right)(a)\\
	    	&=\left(S_{j_0}\sum_{b=0}^{j_1}\binom{j_1}{b}\binom{i}{b}\cdot(b!)\cdot X^{i-b}\left(\sum_{c=0}^{w-1}(b_i(c)\alpha_c^b)S_cD_bf\right)\right)(a)\\	&=\left(S_{j_0}\sum_{b=0}^{j_1}\binom{j_1}{b}\binom{i}{b}\cdot(b!)\cdot X^{i-b}\left(\sum_{c=0}^{w-1}(b_i(c)\alpha_c^b)L_{bu+c}f\right)\right)(a).
	\end{align*}
Notice that the above expression only involves $L_i$s where $i<s$. 

	\cref{lem:general}-\cref{itm:general_lem_deg_pre,itm:general_lemma_G_dist}: follow almost identically to the corresponding items in \cref{thm:AFRS}.
	
    	Thus $\AfFRS_s^{ \beta}[k,n]$ can be efficiently list-decoded
        up to distance $1-\frac{k-1}{rn}-\frac{1}{w}$ with list size $q^{w-1}$.
	By choosing a large enough $w$ and $s$ we can ensure that $1-\frac{k-1}{rn}-\frac{1}{w}>1-\frac{k}{sn}-\epsilon$.
     	\end{proof}

{\small 
  \bibliographystyle{prahladhurl}
  %\bibliography{/Users/prahladh/Dropbox/Documents/academic/papers/jrnl-names-abb,/Users/prahladh/Dropbox/Documents/academic/papers/prahladhbib,/Users/prahladh/Dropbox/Documents/academic/papers/crossref}
  %
  \bibliography{ideal_lin_codes-bib}
}
\end{document}